\documentclass[oldversion]{aa}

\usepackage[english]{babel}
\usepackage[dvips]{graphicx}
\usepackage{latexsym}
\usepackage{amssymb}
\usepackage{mathrsfs}
\usepackage{multirow}
\usepackage{ctable}
\usepackage{afterpage}

\begin{document}

\newcommand{\hours}{^{\mathrm{h}}}
\newcommand{\mins}{^{\mathrm{m}}}
\newcommand{\secs}{^{\mathrm{s}}}
\newcommand{\degs}{^{\circ}}
\newcommand{\jybeam}{Jy\,beam$^\mathrm{-1}\,$}
\newcommand{\mjybeam}{mJy\,beam$^\mathrm{-1}\,$}

\title{Ionospheric Calibration of Low Frequency Radio Interferometric Observations using the Peeling Scheme}
\subtitle{I. Method Description and First Results}

\titlerunning{Ionospheric Calibration of LF Radio Observations I.}

\author{H.T.~Intema\inst{1}
   \and S.~van~der~Tol\inst{2}
   \and W.D.~Cotton\inst{3}
   \and A.S.~Cohen\inst{4}
   \and I.M.~van~Bemmel\inst{1}
   \and H.J.A.~R\"ottgering\inst{1}}

\authorrunning{H.T.~Intema et al.}

\offprints{H.T.~Intema, \\ \email{intema@strw.leidenuniv.nl}}

\institute{Leiden Observatory, Leiden University, P.O.Box 9513, NL-2300 RA, Leiden, The Netherlands
      \and Electr. Engineering, Math. and Computer Science, Delft University of Technology, Delft, The Netherlands
      \and National Radio Astronomy Observatory, Charlottesville, VA, USA
      \and Naval Research Laboratory, Washington DC, USA}

\date{Received 6 October 2008 / Accepted 20 April 2009}

\abstract{Calibration of radio interferometric observations becomes increasingly difficult towards lower frequencies. Below $\sim 300$ MHz, spatially variant refractions and propagation delays of radio waves traveling through the ionosphere cause phase rotations that can vary significantly with time, viewing direction and antenna location. In this article we present a description and first results of SPAM (Source Peeling and Atmospheric Modeling), a new calibration method that attempts to iteratively solve and correct for ionospheric phase errors. To model the ionosphere, we construct a time-variant, 2-dimensional phase screen at fixed height above the Earth's surface. Spatial variations are described by a truncated set of discrete Karhunen-Lo\`eve base functions, optimized for an assumed power-law spectral density of free electrons density fluctuations, and a given configuration of calibrator sources and antenna locations. The model is constrained using antenna-based gain phases from individual self-calibrations on the available bright sources in the field-of-view. Application of SPAM on three test cases, a simulated visibility data set and two selected 74 MHz VLA data sets, yields significant improvements in image background noise (5--75~percent reduction) and source peak fluxes (up to 25~percent increase) as compared to the existing self-calibration and field-based calibration methods, which indicates a significant improvement in ionospheric phase calibration accuracy.}


\maketitle

\section{Introduction}
\label{sec:intro}

Radio waves of cosmic origin are influenced by the Earth's atmosphere before detection at ground level. At low frequencies (LF; $\lesssim 300$~MHz), the dominant effects are refraction, propagation delay and Faraday rotation caused by the ionosphere (e.g., Thompson, Moran \& Swenson \cite{bib:tms2001}; TMS2001 from here on). For a ground-based interferometer (array from here on) observing a LF cosmic source, the ionosphere is the main source of phase errors in the visibilities. Amplitude errors may also arise under severe ionospheric conditions due to diffraction or focussing (e.g., Jacobson \& Erickson \cite{bib:jacobson1992}). 

The ionosphere causes propagation delay differences between array elements, resulting in phase errors in the visibilities. The delay per array element (antenna from here on) depends on the line-of-sight (LoS) through the ionosphere, and therefore on antenna position and viewing direction. The calibration of LF observations requires phase corrections that vary over the field-of-view (FoV) of each antenna. Calibration methods that determine just one phase correction for the full FoV of each antenna (like self-calibration; e.g., see Pearson \& Readhead \cite{bib:pearsonreadhead1984}) are therefore insufficient.

Ionospheric effects on LF interferometric observations have usually been ignored for several reasons: (i) the resolution and sensitivity of the existing arrays were generally too poor to be affected, (ii) existing calibration algorithms (e.g., self-calibration) appeared to give reasonable results most of the time, and (iii) a lack of computing power made the needed calculations prohibitly expensive. During the last 15~years, two large and more sensitive LF arrays have become operational: the VLA at 74~MHz (Kassim et al. \cite{bib:kassim2007}) and the GMRT at 153 and 235~MHz (Swarup \cite{bib:swarup1991}). Observations with these arrays have demonstrated that ionospheric phase errors are one of the main limiting factors for reaching the theoretical image noise level. 

For optimal performance of these and future large arrays with LF capabilities (such as LOFAR, LWA and SKA), it is crucial to use calibration algorithms that can properly model and remove ionospheric contributions from the visibilities. Field-based calibration (Cotton et al.~\cite{bib:cotton2004}) is the single existing ionospheric calibration \& imaging method that incorporates direction-dependent phase calibration. This technique has been succesfully applied to many VLA 74~MHz data sets, but is limited by design for use with relatively compact arrays. 

In Section~\ref{sec:iono_calib}, we discuss ionospheric calibration in more detail. In Section~\ref{sec:method}, we present a detailed description of SPAM, a new ionospheric calibration method that is applicable to LF observations with relatively larger arrays. In Section~\ref{sec:apps}, we present the first results of SPAM calibration on simulated and real VLA 74~MHz observations and compare these with results from self-calibration and field-based calibration. A discussion and conclusions are presented in Section~\ref{sec:concl}.

\section{Ionosphere and Calibration}
\label{sec:iono_calib}

In this Section, we describe some physical properties of the ionosphere, the phase effects on radio interferometric observations and requirements for ionospheric phase calibration.

\subsection{The Ionosphere}
\label{sec:ionosphere}

The ionosphere is a partially ionised layer of gas between $\sim 50$ and $1000$~km altitude over the Earth's surface (e.g., Davies~\cite{bib:davies1990}). It is a dynamic, inhomogeneous medium, with electron density varying as a function of position and time. The state of ionization is mainly influenced by the Sun through photo-ionization at UV and short X-ray wavelengths and through injection of charged particles from the solar wind. Ionization during the day is balanced by recombination at night. The peak of the free electron density is located at a height around 300~km. The free electron column density along a LoS through the ionosphere is generally referred to as \textit{total electron content}, or TEC. The TEC unit (TECU) is $10^{16}$~m$^{-2}$ which is a typically observed value at zenith during nighttime.

The refraction and propagation delay are caused by a varying refractive index $n$ of the ionospheric plasma along the wave trajectory. For a cold, collisionless plasma without magnetic field, $n$ is a function of the free electron density $n^{}_e$ and is defined by (e.g., TMS2001)
\begin{equation}
n^2 = 1 - \frac{ \nu_{p}^{2} }{ {\nu_{}^{2}} },
\label{eq:refractive_index}
\end{equation}
with $\nu$ the radio frequency and $\nu^{}_\mathrm{p}$ the plasma frequency, given by
\begin{equation}
\nu^{}_\mathrm{p} = \frac{ e }{ 2 \pi } \sqrt{ \frac{ n^{}_e }{ \epsilon^{}_0 m } },
\label{eq:plasma_gyro_frequencies}
\end{equation}
with $e$ the electron charge, $m$ the electron mass, $\epsilon^{}_0$ the vacuum permittivity. Typically, for the ionosphere, $\nu^{}_p$ ranges from 1--10~MHz, but may locally rise up to $\sim 200$~MHz in the presence of sporadic E-layers (clouds of unusually high free electron density). Cosmic radio waves with frequencies below the plasma frequency are reflected by the ionosphere and do not reach the Earth's surface. For higher frequencies, the spatial variations in electron density cause local refractions of the wave (Snell's Law) as it travels through the ionosphere, thereby modifying the wave's trajectory. The total propagation delay, integrated along the LoS, results in a phase rotation given by
\begin{equation}
\phi^\mathrm{ion}_{} = - \frac{2 \pi \nu}{c} \int ( n - 1 ) \, \mathrm{d}l,
\label{eq:phase_delay}
\end{equation}
with $c$ the speed of light in vacuum. For frequencies $\nu \gg \nu^{}_p$, this can be approximated by
\begin{equation}
\phi^\mathrm{ion}_{} \approx \frac{ \pi }{ c \nu } \int {\nu^{}_p}^2 dl = \frac{e^2_{}}{4 \pi \epsilon^{}_0 m c \nu} \int n^{}_e \, \mathrm{d}l,
\label{eq:phase_approx}
\end{equation}
where the integral over $n^{}_e$ on the right is the TEC along the LoS. Note that this integral depends on the wave's trajectory, and therefore on local refraction. Because the refractive index is frequency-dependent, the wave's trajectory changes with frequency. As a consequence, the apparent scaling relation $\phi^\mathrm{ion}_{} \propto \nu^{-1}_{}$ from Equation~\ref{eq:phase_approx} is only valid to first order in frequency.

Although bulk changes in the large scale TEC (e.g., a factor of 10 increase during sunrise) have the largest amplitudes, the fluctuations on relatively small spatial scales and short temporal scales are most troublesome for LF interferometric observations. Most prominent are the traveling ionospheric disturbances (TIDs), a response to acoustic-gravity waves in the neutral atmosphere (e.g., van Velthoven \cite{bib:vvelthoven1990}). Typically, medium-scale TIDs are observed at heights between 200 and 400 km, have wavelengths between 250 and 400 km, travel with near-horizontal velocities between 300 and 700 km~h$^{-1}$ in any direction and cause 1--5 percent variations in TEC (TMS2001). 

The physics behind fluctuations on the shortest spatial and temporal scales is less well understood. Temporal and spatial behaviour may be coupled through quasi-frozen patterns that move over the area of interest with a certain velocity and direction (Jacobson \& Erickson \cite{bib:jacobson1992}). Typical variations in TEC are on the order of 0.1 percent, observed on spatial scales of tens of kilometers down to a few km, and time scales of minutes down to a few tens of seconds. The statistical behaviour of radio waves passing through this medium suggests the presence of a turbulent layer with a power-law spectral density of free electron density fluctuations $P^{}_{n^{}_{e}}(q) \propto q^{- \alpha}$ (e.g., TMS2001), with $q \equiv |\vec{q}|$ the magnitude of the 3-dimensional spatial frequency. $P^{}_{n^{}_{e}}(q)$ is defined in units of electron density squared per spatial frequency. The related 2-dimensional structure function of the phase rotation $\phi$ of emerging radio waves from a turbulent ionospheric layer is given by
\begin{equation}
D^{}_{\phi} = \langle [ \phi(\vec{x}) - \phi(\vec{x}+\vec{r}) ]^2 \rangle \propto r^{\gamma}_{},
\label{eq:str_func_body}
\end{equation}
where $\vec{x}$ and $\vec{x} + \vec{r}$ are Earth positions, $r \equiv |\vec{r}|$ is the horizontal distance between these two points, $\langle\dots\rangle$ denotes the expected value and $\gamma = \alpha - 2$. For pure Kolmogorov turbulence, $\alpha = 11 / 3$, therefore $\gamma = 5/3$.

Using differential Doppler-shift measurements of satellite signals, van Velthoven (\cite{bib:vvelthoven1990}) found a power-law relation between spectral amplitude of small-scale ionospheric fluctuations and latitudinal wave-number with exponent $\alpha / 2 = 3 / 2$. Combining with radio interferometric observations of apparent cosmic source shifts, van Velthoven derived a mean height for the ionospheric perturbations of 200--250~km. Through analysis of differential apparent movement of pairs of cosmic sources in the VLSS, Cohen \& R\"ottgering (\cite{bib:cohenrottgering2008}) find typical values for $\gamma / 2$ of 0.50 during nighttime and 0.69 during daytime. Direct measurement of phase structure functions from different GPS satellites (van der Tol, \textit{unpublished}) shows a wide distribution of values for $\gamma$ that peaks at $\sim 1.5$. On average, these results indicate the presence of a turbulent layer below the peak in the free electron density that has more power in the smaller scale fluctuations than in the case of pure Kolmogorov turbulence. Note that for individual observing times and locations, the behaviour of small-scale ionospheric fluctuations may differ significantly from this average.

\subsection{Image Plane Effects}
\label{sec:img_effects}

Interferometry uses the phase differences as measured on baselines to determine the angle of incident waves, and is therefore only sensitive to TEC differences. A baseline is sensitive to TEC fluctuations with linear sizes that are comparable to or smaller than the baseline length. At 75~MHz, a 0.01~TECU difference on a baseline causes a $\sim 1$~radian visibility phase error (Equation~\ref{eq:phase_approx}). Because the observed TEC varies with time, antenna position and viewing direction, visibility phases are distorted by time-varying differential ionospheric phase rotations. 

An instantaneous spatial phase gradient over the array in the direction of a source causes an apparent position shift in the image plane (e.g., Cohen \& R\"ottgering \cite{bib:cohenrottgering2008}), but no source deformation. If the spatial phase behaviour deviates from a gradient, this will also distort the apparent shape of the source. Combining visibilities with different time labels while imaging causes the image plane effects to be time-averaged. A non-zero time average of the phase gradient results in a source shift in the final image. Both a zero-mean time variable phase gradient and higher order phase effects cause smearing and deformation of the source image, and consequently a reduction of the source peak flux (see Cotton \& Condon~\cite{bib:cottoncondon2002} for an example). In the latter case, if the combined phase errors behave like Gaussian random variables, a point source in the resulting image experiences an increase of the source width and reduction of the source peak flux, but the total flux (the integral under the source shape) is conserved. 

For unresolved sources, the \textit{Strehl ratio} is defined as the ratio of observed peak flux over true peak flux. In case of Gaussian random phase errors, the Strehl ratio $R$ is related to the RMS phase error $\sigma^{}_{\phi}$ by (Cotton et al. \cite{bib:cotton2004})
\begin{equation}
R = \exp\left( - \frac{\sigma^{2}_{\phi}}{2} \right).
\label{eq:strehl_ratio}
\end{equation}
A larger peak flux is equivalent to a smaller RMS phase error. This statement is more generally true, because all phase errors cause scattering of source power into sidelobes.

A change in the apparent source shape due to ionospheric phase errors leads to an increase in residual sidelobes after deconvolution. Deconvolution subtracts a time-averaged source image model from the visibility data at all time stamps. In the presence of time-variable phase errors, the mean source model deviates from the apparent, instantaneous sky emission and subtraction is incomplete. Residual sidelobes increase the RMS background noise level and, due to its non-Gaussian character, introduce structure into the image that mimics real sky emission. In LF observations, due to the scaling relation of the dirty beam with frequency (width~$\propto \nu^{-1}_{}$), residual sidelobes around bright sources can be visible at significant distances from the source.

\subsection{Ionospheric Phase Calibration}
\label{sec:calib}

\begin{figure*}
\begin{center}
\resizebox{0.8\hsize}{!}{\includegraphics[angle=0]{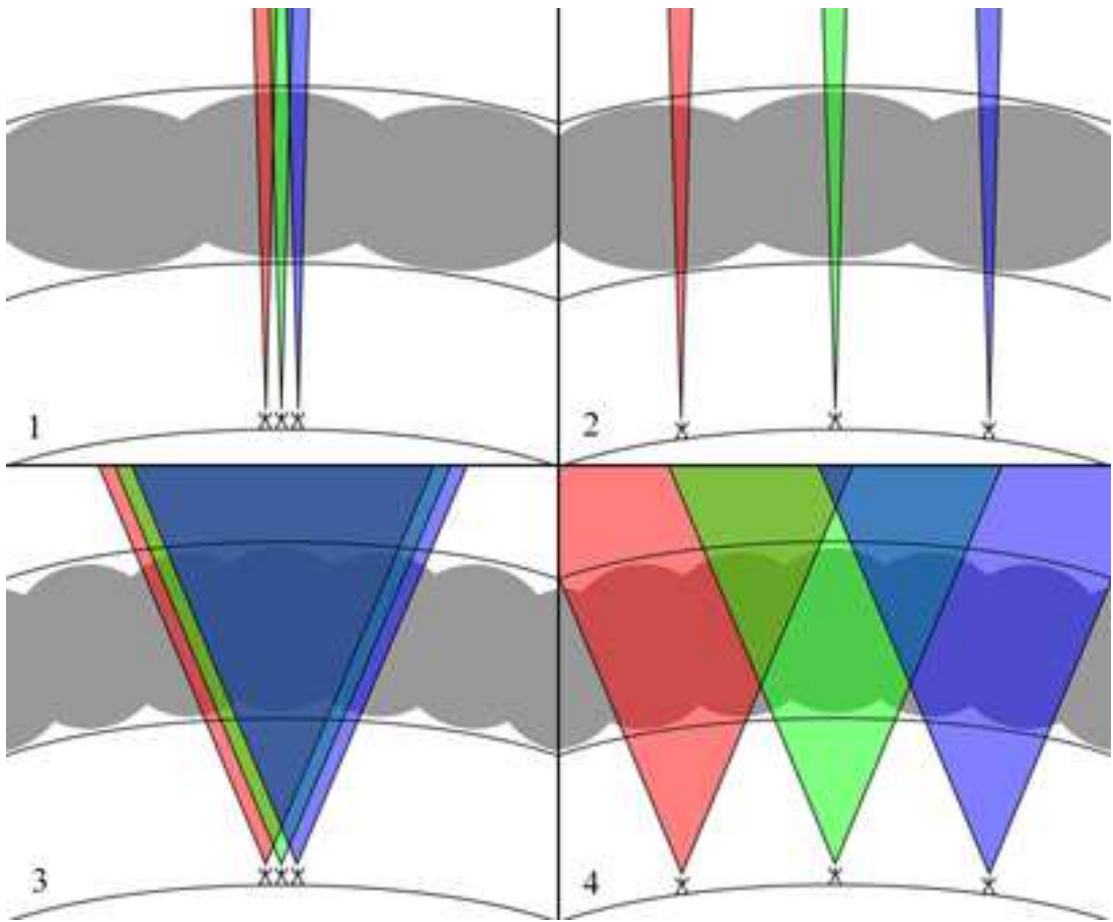}}
\caption{Schematic overview of the different calibration regimes as discussed by Lonsdale (\cite{bib:lonsdale2005}). For clarity, only two spatial dimensions and one calibration time interval are considered. In this overview, the array is represented by three antennas at ground level, looking through the ionospheric electron density structure (grey bubbles) with individual fields-of-view (red, green and blue areas). Due to the relatively narrow primary beam patterns in regimes 1 and 2 (\textit{top left} and \textit{top right}, respectively), each individual antenna 'sees' an approximately constant TEC across the FoV. The relatively wide primary beam patterns in regimes 3 and 4 (\textit{bottom left} and \textit{bottom right}, respectively) causes the antennas to 'see' TEC variations across the FoV. For the relatively compact array configurations in regimes 1 and 3, the TEC variation across the array for a single viewing direction within the FoV is approximately a gradient. For the relatively extended array configurations in regimes 2 and 4, the TEC variation across the array for a single viewing direction differs significantly from a gradient. The consequences for calibration of the array are discussed in the text.}
\label{fig:regimes}
\end{center}
\end{figure*}

Lonsdale (\cite{bib:lonsdale2005}) discussed four different regimes for (instantaneous) ionospheric phase calibration, depending on the different linear spatial scales involved. These scales are the array size $A$, the scale size $S$ of ionospheric phase fluctuations and the projected size $V$ of the field-of-view (FoV) at a typical ionospheric height. We use the term \textit{compact} array when $A \ll S$ and \textit{extended} array when $A \gtrsim S$. Note that these definitions change with ionospheric conditions, so there is no fixed linear scale that defines the difference between compact and extended. A schematic overview of the different regimes is given in Figure~\ref{fig:regimes}.

The combination $A V / S^{2}_{}$ is a measure of the complexity of ionospheric phase calibration. Both $S$ and $V$ depend on the observing frequency $\nu$. For a power-law spectral density of free electron density fluctuations (see Section~\ref{sec:ionosphere}) $S$ scales with $\nu$ , and for a fixed circular antenna aperture $V$ scales with $\nu^{-1}_{}$. Therefore, $A V / S^{2}_{}$ scales with $\nu^{-3}_{}$, signalling a rapid increase in calibration problems towards low frequencies.

Under \textit{isoplanatic} conditions ($V \ll S$), the ionospheric phase error per antenna does not vary with viewing direction within the FoV, for both compact and large arrays (Lonsdale regimes 1 and 2, respectively). Phase-only self-calibration on short enough time-scales is sufficient to remove the ionospheric phase errors from the visibilities. 

Under \textit{anisoplanatic} conditions ($V \gtrsim S$), the ionospheric phase error varies over the FoV of each antenna. A single phase correction per antenna is no longer sufficient. Self-calibration may still converge, but the resulting phase correction per antenna is a flux-weighted average of ionospheric phases across the FoV (see Section~\ref{sec:instr_calib}). Accurate self-calibration and imaging of individual very bright and relatively compact sources is therefore possible, even with extended arrays (see Gizani, Cohen \& Kassim \cite{bib:gizani2005} for an example). For a compact array (Lonsdale regime~3), the FoV of different antennas effectively overlap at ionospheric height. The LoS of different antennas towards one source run close and parallel through the ionosphere. For an extended array (Lonsdale regime~4), the FoV of different antennas may partially overlap at ionospheric height, but not necessarily. Individual LoS from widespread antennas to one source may trace very different paths through the ionosphere

In regime~3, ionospheric phases behave as a spatial gradient over the array that varies with viewing direction. This causes the apparent position of sources to change with time and viewing direction, but no source deformation takes place. The 3-dimensional phase structure of the ionosphere can be effectively reduced to a 2-dimensional phase screen, by integrating the free electron density along the LoS (Equation~\ref{eq:phase_approx}). Radio waves that pass the virtual screen experience an instantaneous ionospheric phase rotation depending on the \textit{pierce point} position (where the LoS pierces the phase screen). When assuming a fixed number of required ionospheric parameters per unit area of phase screen, calibration of a compact array requires a minimal number of parameters because each antenna illuminates the same part of the phase screen. 

In regime~4, the dependence of ionospheric phase on antenna position and viewing direction is more complex. This causes source position shifts and source shape deformations that both vary with time and viewing direction. A 2-dimensional phase screen model may still be used, but only when the dominant phase fluctuations originate from a restricted height range $\Delta h \ll S$ in the ionosphere. The concept of a thin layer at a given height is attractive, because it reduces the complexity of the calibration problem drastically. When using an airmass function to incorporate a zenith angle dependence, the spatial phase function is in effect reduced to 2~spatial dimensions. Generally, a phase screen in regime~4 requires a larger number of model parameters than in regime~3, because the phase screen area illuminated by the total array is larger. 

It is currently unclear under which conditions a 2-dimensional phase screen model becomes too inaccurate to model the ionosphere in regime~4. For very long baselines or very severe ionospheric conditions, a full 3-dimensional ionospheric phase model may be required, where ionospheric phase corrections need to determined by ray-tracing. Such a model is likely to require many more parameters than can be extracted from radio observations alone. To first order, it may be sufficient to extend the phase screen model with some form of height-dependence. Examples of such extensions are the use of several phase screens at different heights (Anderson~\cite{bib:anderson2006}) or introducing smoothly varying partial derivatives of TEC or phase as a function of zenith angle (Noordam et al.~\cite{bib:noordam2008}). 

Calibration needs to determine corrections on sufficiently short time scales to track the ionospheric phase changes. The phase rate of change depends on the intrinsic time variability of the TEC along a given LoS and on the speed of the LoS from the array antennas through the ionosphere while tracking a cosmic source. The latter may range up to $\sim 100$~km~h$^\mathrm{-1}$ at 200~km height. The exact requirements on the time resolution of the calibration are yet to be determined. In principle, the time-variable ionospheric phase distortions needs to be sampled at least at the Nyquist frequency. However, during phase variations of large amplitude ($\gg 1$~radian), $2 \pi$ radian phase winding introduces periodicity on much shorter time scales. To succesfully unwrap phase winds, at least two corrections per $2 \pi$ radian phase change are required.

\subsection{Proposed and Existing Ionospheric Calibration Schemes}
\label{sec:existing}

Schwab (\cite{bib:schwab1984}) and Subrahmanya (\cite{bib:subrahmanya1991}) have proposed modifications to the self-calibration algorithm to support direction-dependent phase calibration. Both methods discuss the use of a spatial grid of interpolation nodes (additional free parameters) to characterize the spatial variability of the ionospheric phase error. Schwab suggests to use a different set of nodes per antenna, while Subrahmanya suggests to combine these sets by positioning them in a quasi-physical layer at fixed height above the Earth's surface (this to reduce the number of required nodes when the FoVs from different antennas overlap at ionospheric height). Neither of both proposed methods have been implemented.

Designed to operate in Lonsdale regime 3, field-based calibration by Cotton et al. (\cite{bib:cotton2004}) is the single existing implementation of a direction-dependent ionospheric phase calibration algorithm. Typically, for each time interval of 1--2 minutes of VLA 74~MHz data, the method measures and converts the apparent position shift of 5--10 detectable bright sources within the FoV into ionospheric phase gradients over the array. To predict phase gradients in arbitrary viewing directions for imaging of the full FoV, an independent phase screen per time interval is fitted to the measured phase gradients. The phase screen is described by a 5~term basis of Zernike polynomials (up to second order, excluding the constant zero order).

Field-based calibration has been used to calibrate 74~MHz VLA observations, mostly in B-configuration (e.g., Cohen et al. \cite{bib:cohen2007}) but also several in A-configuration (e.g., Cohen et al. \cite{bib:cohen2003}, \cite{bib:cohen2004}). Image plane comparison of field-based calibration against self-calibration shows an overall increase of source peak fluxes (in some cases up to a factor of two) and reduction of residual sidelobes around bright sources, a clear indication of improved phase calibration over the FoV (Cotton \& Condon~\cite{bib:cottoncondon2002}). The improved overall calibration performance sometimes compromises the calibration towards the brightest source. 

Zernike polynomials are often used to describe abberations in optical systems, because lower order terms match well with several different types of wavefront distortions, and the functions are an orthogonal set on the circular domain of the telescope pupil. Using Zernike polynomials to describe an ionospheric phase screen may be less suitable, because they are not orthogonal on the discrete domain of pierce points, diverge when moving away from the field center and have no relation to ionospheric image abberations (except for first order, which can model a large scale TEC gradient). Non-orthogonality leads to interdependence between model parameters, while divergence is clearly non-physical and leads to undesirable extrapolation properties.

For extended LF arrays or more severe ionospheric conditions, the ionospheric phase behaviour over the array for a given viewing direction is no longer a simple gradient. Under these conditions, performance of field-based calibration degrades. For the 74~MHz VLA Low-frequency Sky Survey (VLSS; Cohen et al. \cite{bib:cohen2007}), field-based calibration was unable to calibrate the VLA in B-configuration for about 10--20\% of the observing time due to severe ionospheric conditions. Observing at 74~MHz with the $\sim 3$ times larger VLA A-configuration leads to a relative increase in the failure rate of field-based calibration. This is to be expected, as the larger array size results in an increased probability for the observations to reside in Lonsdale regime~4.

The presence of higher order phase structure over the array in the direction of a calibrator requires an antenna-based phase calibration rather than a source position shift to measure ionospheric phases. The calibration methods proposed by Schwab and Subrahmanya (see above) do allow for higher order phase corrections over the array and could, in principle, handle more severe ionospheric conditions. An alternative approach is to use the \textit{peeling} technique (Noordam~\cite{bib:noordam2004}), which consist of sequential self-calibrations on individual bright sources in the FoV. This yields per source a set of time-variable antenna-based phase corrections and a source model. Because the peeling corrections are applicable to a limited set of viewing directions, they need to be interpolated in some intelligent way to arbitrary viewing directions while imaging the full FoV. Peeling is described in more detail in Section~\ref{sec:peeling}.

Noordam (\cite{bib:noordam2004}) has proposed a `generalized' self-calibration method for LOFAR (e.g., R\"ottgering et al.~\cite{bib:rottgering2006}) that includes calibration of higher order ionospheric phase distortions. Similar to `classical' self-calibration, instrumental and environmental (including ionospheric) parameters are estimated by calibration against a sky brightness model. Sky model and calibration parameters are iteratively updated to converge to some final result. Uniqueness of the calibration solution is controlled by putting restrictions on the time-, space- and frequency behaviour of the fitted parameters. The effects of the ionosphere are modeled in a Minimum Ionospheric Model (MIM; Noordam~\cite{bib:noordam2008}), which is yet to be defined in detail. The philosophy of the MIM is to use a minimal number of physical assumptions and free parameters to accurately reproduce the observed effects of the ionosphere on the visibilities for a wide-as-possible range of ionospheric conditions. The initial MIM is to be constrained using peeling corrections.

\section{Method}
\label{sec:method}

SPAM, an abbreviation of `Source Peeling and Atmospheric Modeling', is the implementation of a new ionospheric calibration method, combining several concepts from proposed and existing calibration methods. SPAM is designed to operate in Lonsdale regime 4 and can therefore also operate in regimes 1 to 3. It uses the calibration phases from peeling sources in the FoV to constrain an ionospheric phase screen model. The phase screen mimics a thin turbulent layer at a fixed height above the Earth's surface, in concordance with the observations of ionospheric small-scale structure (Section~\ref{sec:ionosphere}). The main motivation for this work was to test several aspects of ionospheric calibration on existing VLA and GMRT data sets on viability and qualitative performance, and thereby support the development of more advanced calibration algorithms for future instruments such as LOFAR.

Generally, the instantaneous ionosphere can only be sparsely sampled, due to the non-uniform sky distribution of a limited number of suitable calibrators and an array layout that is optimized for UV-coverage rather than ionospheric calibration. To minimize the error while interpolating to unsampled regions, an optimal choice of base functions for the description of the phase screen is of great importance. Based on the work by van der Tol \& van der Veen (\cite{bib:vdtolvdveen2007}), we use the discrete Karhunen-Lo\`eve (KL) transform to determine an optimal set of base `functions' to describe our phase screen. For a given pierce point layout and an assumed power-law slope for the spatial structure function of ionospheric phase fluctuations (see Section~\ref{sec:ionosphere}), the KL transform yields a set of base vectors with several important properties: (i) the vectors are orthogonal on the pierce point domain, (ii) truncation of the set (reduction of the model order) gives a minimal loss of information, (iii) interpolation to arbitrary pierce point locations obeys the phase structure function, and (iv) spatial phase variability scales with pierce point density, i.e., most phase screen structure is present in the vicinity of pierce points, while it converges to zero at infinite distance. More detail on this phase screen model is given in Section~\ref{sec:iono_model}.

Because the required calibration time resolution is still an open issue, and the SPAM model does not incorporate any restrictions on temporal behaviour, independent phase screens are determined at the highest possible time resolution (which is the visibility integration time resolution). 

SPAM calibration can be separated in a number of functional steps, each of which is discussed in detail in the sections to follow. The required input is a spectral-mode visibility data set that has flux calibration and bandpass calibration applied, and radio frequency interference (RFI) excised (see Lazio, Kassim \& Perley~\cite{bib:lazio2005} or Cohen et al.~\cite{bib:cohen2007} for details). The SPAM recipe consists of the following steps:
\begin{enumerate}
\item{Obtain and apply instrumental calibration corrections for phase (Section~\ref{sec:instr_calib}).}
\label{enu:spam_instr_calib}
\item{Obtain an initial model of the apparent sky, together with an initial ionospheric phase calibration (Section~\ref{sec:init_sky_calib}).}
\label{enu:spam_init_sky_calib}
\item{Subtract the sky model from the visibility data while applying the phase calibration. Peel apparently bright sources (Section~\ref{sec:peeling}).}
\label{enu:spam_peeling}
\item{Fit an ionospheric phase screen model to the peeling solutions (Section~\ref{sec:iono_model}).}
\label{enu:spam_iono_mnodel}
\item{Apply the model phases on a facet-to-facet basis during re-imaging of the apparent sky (Section~\ref{sec:imaging}).}
\label{enu:spam_imaging}
\end{enumerate}
Steps~\ref{enu:spam_peeling} to \ref{enu:spam_imaging} define the SPAM calibration cycle, as the image produced in step~\ref{enu:spam_imaging} can serve as an improved model of the apparent sky in step~\ref{enu:spam_peeling}.

The scope of applications for SPAM is limited by a number of assumptions that were made to simplify the current implementation:
\begin{itemize}
\item{The ionospheric inhomogeneities that cause significant phase distortions are located in a single, relatively narrow height range.}
\item{There exists a finitely small angular patch size, which can be much smaller than the FoV of an individual antenna, over which the ionospheric phase contribution is effectively constant. Moving from one patch to neighbouring patches results in small phase transitions ($\ll 1$~radian).}
\item{There exists a finitely small time range, larger than the integration time interval of an observation, over which the apparent ionospheric phase change for any of the array antennas along any line-of-sight is much smaller than a radian.}
\item{The bandwidth of the observations is small enough to be effectively monochromatic, so that the ionospheric dispersion of waves within the frequency band is negligible.}
\item{Within the given limitations on bandwidth and integration time, the array is sensitive enough to detect at least a few ($\gtrsim 5$) sources within the target FoV that may serve as phase calibrators.}
\item{The ionospheric conditions during the observing run are such that self-calibration is able to produce a good enough initial calibration and sky model to allow for peeling of multiple sources. This might not work under very bad ionospheric conditions, but for the applications presented in this article it proved to be sufficient.}
\item{After each calibration cycle (steps~\ref{enu:spam_peeling} to \ref{enu:spam_imaging}), the calibration and sky model are equally or more accurate than the previous. This implies convergence to a best achievable image.}
\item{The instrumental amplitude and phase contributions to the visibilities, including the antenna power patterns projected onto the sky towards the target source, are constant over the duration of the observing run.}
\end{itemize}
SPAM does not attempt to model the effects of ionospheric Faraday rotation on polarization products, and is therefore only applicable to intensity measurements (stokes I). 

In our implementation we have focussed on functionality rather than processing speed. In its current form, SPAM is capable of processing quite large offline data sets, but is not suitable for real-time processing as is required for LOFAR calibration. SPAM relies heavily on functionality available in NRAO's Astronomical Image Processing System (AIPS; e.g., Bridle \& Greisen \cite{bib:bridlegreisen1994}). It consists of a collection of Python scripts that accesses AIPS tasks, files and tables using the ParselTongue interface (Kettenis et al. \cite{bib:kettenis2006}). Two main reasons to use AIPS are its familiarity and proven robustness while serving a large group of users over a 30~year lifetime, and the quite natural way by which the ionospheric calibration method is combined with polyhedron imaging (Perley \cite{bib:perley1989a}; Cornwell \& Perley \cite{bib:cornwellperley1992}). SPAM uses a number of 3$^\mathrm{rd}$ party Python libraries, like scipy, numpy and matplotlib for math and matrix operations and plotting. For non-linear least squares fitting of ionospheric phase models, we have adopted a Levenberg-Marquardt solver (LM; e.g., Press et al. \cite{bib:press1992}) based on IDL's MPFIT package (Markwardt~\cite{bib:markwardt2008}).

\subsection{Instrumental Phase Calibration}
\label{sec:instr_calib}

Each antenna in the array adds an instrumental phase offset to the recorded signal before correlation. At low frequencies, changes in the instrumental signal path length (e.g., due to temperature induced cable length differences) are very small compared to the wavelength, therefore instrumental phase offsets are generally stable over long time periods (hours to days). SPAM requires removal of the instrumental phase offsets from the visibilities prior to ionospheric calibration. 

Instead of directly measuring the sky intensity $I(l,m)$ as a function of viewing direction cosines $(l,m)$, an interferometer measures an approximate Fourier transform of the sky intensity. For a baseline consisting of antennas $i$ and $j$, the perfect response to all visible sky emission for a single time instance and frequency is given by the measurement equation (ME) for visibilities (e.g., TMS2001):
\begin{equation}
V^{}_{ij} = \int \int I(l,m) e^{- 2 \pi J \left[ u^{}_{ij} l + v^{}_{ij} m + w^{}_{ij} ( n - 1 ) \right]} \frac{\mathrm{d}l \, \mathrm{d}m}{n},
\label{eq:meas_eq}
\end{equation}
where $J$ indicates the imaginary part of a complex number, $n = \sqrt{1 - l^{2}_{} - m^{2}_{}}$, $u^{}_{ij}$ and $v^{}_{ij}$ are baseline coordinates in the UV plane (expressed in wavelengths) parallel to $l$ and $m$, respectively, and $w^{}_{ij}$ is the perpendicular baseline coordinate along the LoS towards the chosen celestial \textit{phase tracking center} at $(l,m) = (0,0)$. In practise, these measurements are modified with predominantly antenna-based complex gain factors $a^{}_{i}$ that may vary with time, frequency, antenna position and viewing direction. This modifies the ME into
\begin{eqnarray}
\hat{V}^{}_{ij} & = & \int \int a^{}_{i}(l,m) \, a^{\dagger}_{j}(l,m) \nonumber \\
 & & \quad I(l,m) e^{- 2 \pi J \left[ u^{}_{ij} l + v^{}_{ij} m + w^{}_{ij} ( n - 1 ) \right]} \frac{\mathrm{d}l \, \mathrm{d}m}{n}.
\label{eq:mod_meas_eq}
\end{eqnarray}
Determination of the gain factors is generally referred to as \textit{calibration}. When known, only gain factors that do not depend on viewing direction can be removed from the visibility data prior to image reconstruction by applying the calibration:
\begin{equation}
V^{}_{ij} = ( a^{}_{i} \, a^{\dagger}_{j} )^{-1}_{} \hat{V}^{}_{ij}
\label{eq:calibration}
\end{equation}
This operation is generally not possible for gain factors that do depend on viewing direction, because these gain factors cannot be moved in front of the integral in Equation~\ref{eq:mod_meas_eq}. One may still choose to apply gain corrections for a single viewing direction (e.g. to image a particular source), but the accuracy of imaging and deconvolution of other visible sources will degrade when moving away from the selected viewing direction. A solution for wide-field imaging and deconvolving in the presence of direction-dependent gain factors is discussed in Section~\ref{sec:imaging}.

The standard approach for instrumental phase calibration at higher frequencies is to repeatedly observe a bright (mostly unresolved) source during an observing run. Antenna-based gain phase corrections $g^{}_{i} \approx a^{-1}_{i}$ are estimated by minimizing the weighted difference sum $S$ between observed visibilities $\hat{V}^{}_{ij}$ and source model visibilities $V^\mathrm{model}_{ij} \approx V^{}_{ij}$ (e.g., TMS2001; implemented in AIPS task CALIB):
\begin{equation}
S = \sum^{}_{i} \sum^{}_{j>i} W^{}_{ij} \| V^\mathrm{model}_{ij} - g^{}_{i} \, g^{\dagger}_{j} \hat{V}^{}_{ij} \|^{p}_{},
\label{eq:calibration2}
\end{equation}
with $W^{}_{ij}$ the visibility weight (reciproke of the uncertainty in the visibility measurement), $g^{}_{i} = e^{ i \phi^\mathrm{cal}_{i} }_{}$ and $p$ the power of the norm (typically 1 or 2). The source model visibilities $V^\mathrm{model}_{ij}$ are calculated using Equation~\ref{eq:meas_eq} with $I(l,m) = I^\mathrm{model}_{}(l,m)$. The phase corrections $\phi^\mathrm{cal}_{i}$ consist of an instrumental and an atmospheric part. The corrections are interpolated in time and applied to the target field visibilities, under the assumptions that the instrumental and atmospheric phase offsets vary slowly in time, and that the atmospheric phase offsets in the direction of the target are equal to those in the direction of the calibrator. 

At low frequencies, there are two complicating factors for the standard approach: (i) the FoV around the calibrator source is large and includes many other sources, and (ii) the ionospheric phase offset per antenna changes significantly with time and viewing direction. The former can be overcome by choosing a very bright calibrator source with a flux that dominates over the combined flux of all other visible sources on all baselines. For the VLSS (Cohen et al. \cite{bib:cohen2007}), the 17,000~Jy of Cygnus A was more than sufficient to dominate over the total apparent flux of $400 - 500$~Jy in a typical VLSS field. The latter requires filtering of the phase corrections to extract only the instrumental part, which is then applied to the target field visibilities.

For SPAM, we have adopted an instrumental phase calibration method that is very similar to the procedure used for field-based calibration (Cotton et al. \cite{bib:cotton2004}). Antenna-based phase corrections are obtained on the highest possible time resolution by calibration on a very bright source $k$ using the robust L1 norm (Equation~\ref{eq:calibration2} with $p = 1$; Schwab~\cite{bib:schwab1981}). A phase correction $\phi^\mathrm{cal}_{ikn}$ for antenna $i$ at time interval $n$ consist of several contributions:
\begin{equation}
\phi^\mathrm{cal}_{ikn} = \phi^\mathrm{instr}_{i} + \phi^\mathrm{ion}_{ikn} - \phi^{}_{rkn} - \phi^\mathrm{ambig}_{ikn},
\label{eq:gain_phase}
\end{equation}
where the instrumental and ionospheric phase corrections, $\phi^\mathrm{instr}_{i}$ and $\phi^\mathrm{ion}_{ikn}$ respectively, are assumed to be constant resp. vary with time and antenna position over the observing run. The other right-hand terms are the phase offset $\phi^{}_{rkn} = \phi^\mathrm{instr}_{r} + \phi^\mathrm{ion}_{rkn}$ of an arbitrarily chosen reference antenna $r \in \{i\}$, and the phase ambiguity term $\phi^\mathrm{ambig}_{ikn} = 2 \pi N^{}_{ikn}$ with integer $N^{}_{ikn}$ that maps $\phi^\mathrm{cal}_{ikn}$ into the $[0,2\pi)$ domain.

The antenna-based phase corrections are split into instrumental and ionospheric parts on the basis of their temporal and spatial behaviour. The phase corrections are filtered by iterative estimation of invariant instrumental phases (together with the phase ambiguities) and time- and space-variant ionospheric phases. The instrumental phases are estimated by robust averaging ($+3\,\sigma$ rejection) over all time intervals $n$:
\begin{equation}
\tilde{\phi}^\mathrm{instr}_{i} = \left< \left( \phi^\mathrm{cal}_{ikn} - \tilde{\phi}^\mathrm{ion}_{i} \right) \bmod{ 2 \pi } \right>_n.
\label{eq:instr_est}
\end{equation}
The phase ambiguity estimates follow from
\begin{equation}
\tilde{\phi}^\mathrm{ambig}_{ikn} = 2 \pi \, \mathrm{round}\left( \left[ \tilde{\phi}^\mathrm{instr}_{i} + \tilde{\phi}^\mathrm{ion}_{i} - \phi^\mathrm{cal}_{ikn} \right] / 2 \pi \right),
\label{eq:ambig_est}
\end{equation}
where the $\mathrm{round}()$ operator rounds a number to the nearest integer value. The instrumental phase offset of the reference antenna is arbitrarily set to zero. The ionospheric phases are constrained by fitting a time-varying spatial gradient $\vec{G}^{}_{kn}$ to the phases over the array. The gradient fit consists of an initial estimate directly from the calibration phase corrections, followed by a refined fit by using the LM solver to minimize
\begin{eqnarray}
\chi^{2}_{kn} & = & \sum^{}_{i} \bigg[ \left( \phi^\mathrm{cal}_{ikn} - \tilde{\phi}^\mathrm{instr}_{i} + \tilde{\phi}^\mathrm{ambig}_{ikn} \right) - \nonumber \\
& & \quad \quad \quad \quad \underbrace{ \vec{G}^{}_{kn} \cdot \left( \vec{x}^{}_{i} - \vec{x}^{}_{r} \right) }_{ \tilde{\phi}^\mathrm{ion}_{ikn} } \bigg]^{2}_{},
\label{eq:ion_est}
\end{eqnarray}
where $\vec{x}^{}_{i}$ is the position of antenna $i$. The ionospheric phase offset of the reference antenna is arbitrarily set to zero, which makes it a pivot point over which the phase gradient rotates. Higher order ionospheric effects are assumed to average to zero in Equation~\ref{eq:instr_est}.

\subsection{Initial Phase Calibration and Initial Sky Model}
\label{sec:init_sky_calib}

The instrumental phase calibration method described in Section~\ref{sec:instr_calib} assumes that the time-averaged ionospheric phase gradient over the array in the direction of the bright phase calibrator is zero. Any non-zero average is absorbed into the instrumental phase estimates, causing a position shift of the whole target field and thereby invalidating the astrometry. Before entering the calibration cycle (Sections~\ref{sec:peeling}--\ref{sec:imaging}), SPAM requires restoration of the astrometry and determination of an initial sky model and initial ionospheric calibration.

To restore the astrometry, the instrumentally corrected target field data from Section~\ref{sec:instr_calib} is phase calibrated against an apparent sky model (AIPS task CALIB). The default is a point source model, using NVSS catalog positions (Condon et al.~\cite{bib:condon1994}, \cite{bib:condon1998}), power-law interpolated fluxes from NVSS and WENSS/WISH catalogs (Rengelink et al.~\cite{bib:rengelink1997}) and a given primary beam model. To preserve the instrumental phase calibration as obtained in Section~\ref{sec:instr_calib} during further processing, time-variable phase corrections resulting from calibration steps in this and the following sections are stored in a table (AIPS SN table) rather than applied directly to the visibility data. The sky model calibration is followed by wide-field imaging (AIPS task IMAGR) and several rounds of phase-only self-calibration (CALIB and IMAGR) at the highest possible time resolution, yielding the initial sky model and initial phase calibration.

For wide-field imaging with non-coplanar arrays, the standard imaging assumptions that the relevant sky area is approximately flat and the third baseline coordinate (\textit{w}-term in Equation~\ref{eq:meas_eq}) is constant across the FoV are no longer valid. To overcome this, SPAM uses the polyhedron method (Perley \cite{bib:perley1989a}; Cornwell \& Perley \cite{bib:cornwellperley1992}) that divides the large FoV into a hexagonal grid of small, partially overlapping \textit{facets} that individually do satisfy the assumptions above (AIPS task SETFC). Additional facets are centered on relatively bright sources inside and outside the primary beam area to reduce image artefacts due to pixellation (Perley \cite{bib:perley1989b}; Briggs \& Cornwell \cite{bib:briggscornwell1992}; Briggs \cite{bib:briggs1995}; Voronkov \& Wieringa \cite{bib:voronkovwieringa2004}; Cotton \& Uson \cite{bib:cottonuson2007}).

The Cotton-Schwab algorithm (Schwab \cite{bib:schwab1984}; Cotton \cite{bib:cotton1999}; Cornwell, Braun \& Briggs \cite{bib:cornwellbraunbriggs1999}) is a variant of CLEAN deconvolution (H\"ogbom \cite{bib:hogbom1974}; Clark \cite{bib:clark1980}) that allows for simultaneous deconvolution of multiple facets, using a different dirty beam for each facet. \textit{Boxes} are used to restrict CLEANing to real sky emission, making sure that sources are deconvolved in the nearest facet only (CLEAN model components are stored in facet-based AIPS CC tables). After deconvolution, the CLEAN model is restored to the relevant residual facets (AIPS task CCRES) using a CLEAN beam, and the facets are combined to form a single image of the full FoV (AIPS task FLATN).

\subsection{Peeling}
\label{sec:peeling}

To construct a model of ionospheric phase rotations in arbitrary viewing directions within the FoV, SPAM requires measurements in as many directions as possible. When no external sources of ionospheric information are available, the target field visibilities themselves need to be utilized. Calibration on individual bright sources in the FoV can supply the required information, even in the presence of higher order phase structure over the array. After instrumental phase offsets are removed, phase calibration corrections are an relative measure of ionospheric phase:
\begin{equation}
\phi^\mathrm{cal}_{ikn} = \phi^\mathrm{ion}_{ikn} - \phi^\mathrm{ion}_{rkn} - \phi^\mathrm{ambig}_{ikn},
\label{eq:peel_phase}
\end{equation}
where we used Equation~\ref{eq:gain_phase} with $\phi^\mathrm{instr}_{i} = \phi^\mathrm{instr}_{r} = 0$.

SPAM uses the peeling technique (Noordam~\cite{bib:noordam2004}) to obtain phase corrections in different viewing directions. Peeling consists of self-calibration on individual sources, yielding per source a set of time-variable antenna-based phase corrections and a source model, after which the source model is subtracted from the visibility data set while temporarily applying the phase corrections (AIPS tasks SPLIT, UVSUB and CLINV/SPLIT).

For peeling to converge, the source needs to be the dominant contributor of flux to the visibilities on all baselines. Especially at low frequencies, the presence of many other sources in the large FoV adds considerable noise to the peeling phase corrections. To suppress this effect, the following steps are performed: (i) The best available model of the apparent sky is subtracted from the visibility data while temporarily applying the associated phase calibration(s). The initial best available model and associated phase calibration is the self-calibration output of Section~\ref{sec:init_sky_calib}. Individual source models are added back before peeling. (ii) Sources are peeled in decreasing flux order to suppress the effect of brighter sources on the peeling of fainter sources. (iii) Calibration only uses visibilities with projected baseline lengths longer than a certain threshold. This excludes the high `noise' in the visibilities near zero-length baselines from the coherent flux contribution of imperfectly subtracted sources. 

The radio sky can be approximated by a discrete number of isolated, invariant sources of finite angular extend. Visibilities in the ME (Equation~\ref{eq:meas_eq}) for a single integration time $n$ can therefore be split into a linear combination of contributions from individual sources $k$:
\begin{eqnarray}
V^{}_{ijn} & = & \sum^{}_{k} V^{}_{ijkn} = \sum^{}_{k} \int \int I^{}_{k}(l,m) \nonumber \\
 & & \quad e^{- 2 \pi J \left[ u^{}_{ijn} l + v^{}_{ijn} m + w^{}_{ijn} ( n - 1 ) \right]} \frac{\mathrm{d}l \, \mathrm{d}m}{n}.
\label{eq:meas_eq_kn}
\end{eqnarray}
The subtraction of all but the peeling source $k'$ from the measured visibilities in step (i) above can be described as
\begin{equation}
\hat{V}^{}_{ijk'n} \approx \hat{V}^{}_{ijn} - \sum^{}_{k \neq k'} ( g^{}_{ikn} \, g^{\dagger}_{jkn} )^{-1}_{} V^\mathrm{model}_{ijkn},
\label{eq:subtraction}
\end{equation}
with $g^{}_{ikn} = g^{}_{i}( l^{}_{k}, m^{}_{k}, t^{}_{n} ) = e^{i\phi^\mathrm{cal}_{ikn}}_{}$ the best available calibration in the viewing direction of source $k$, and $V^\mathrm{model}_{ijkn}$ the visibilities that are derived from the best available model $I^\mathrm{model}_{ijk}$ of source $k$. The peeling itself consists of iterative calibration and imaging steps of the peeling source $k'$. The calibration (Equation~\ref{eq:calibration2} with $p = 1$) updates the antenna gain corrections $g^{}_{ikn}$ by minimizing
\begin{equation}
S^{}_{n} = \sum^{}_{i} \sum^{}_{j>i} w^{}_{ijn} \| V^\mathrm{model}_{ijk'n} - g^{}_{in} \, g^{\dagger}_{jn} \hat{V}^{}_{ijk'n} \|,
\label{eq:calibration3}
\end{equation}
while the imaging step updates $I^\mathrm{model}_{ijk'}$ and therefore $V^\mathrm{model}_{ijk'n}$.

In practise, due to incompleteness of the sky model and inaccuracies in the phase calibration, there will always remain some contaminating source flux in the visibilities while peeling. Complemented with system noise, sky noise, residual RFI and other possible sources of noise, the noise in the visibilities propagates into the phase corrections from the peeling process. 

Absolute astrometry is not conserved during peeling, because self-calibration allows antenna-based phase corrections to vary without constraint. In subsequent peeling cycles, small non-zero phase gradients in the phase residuals after calibration can cause the source model to wander away from its true position. In SPAM, astrometry errors are minimized by re-centering the source model to its true (catalog) position before calibration in each self-calibration loop. By default, SPAM re-centers the peak of the model flux to the nearest bright point source position in the NVSS catalog (Condon et al. \cite{bib:condon1994}; \cite{bib:condon1998}). It is recommended to visually check the final peeling source images for possible mismatches with the catalog (e.g., in case of double sources or sources with a spatially varying spectral index).

While peeling, SPAM attempts to calibrate sources on the highest possible time resolution, which is the visibility time grid. The noise in the resulting phase corrections depends on the signal-to-noise ratio (SNR) of the source flux in the visibilities. To increase the number of peeling sources and limit the phase noise in case of insufficient SNR, SPAM is allowed to increase the calibration time interval beyond the visibility integration time up to an arbitrary limit. Through image plane analysis, SPAM estimates the required calibration time-interval per source:
\begin{equation}
n^{}_t = \left( \frac{ \sigma^{}_\mathrm{L} }{ \alpha S^{}_{p} } \right)^2_{} N^{}_t,
\label{eq:peel_time_limit}
\end{equation}
where $n^{}_t$ is the required number of integration times in a calibration interval, $N^{}_t$ is the total number of integration times within the observation, $\alpha$ is the minimum required SNR per integration time (a tweakable parameter that sets the balance between the SNR and the time resolution of the peeling phase corrections), and $S^{}_{p}$ and $\sigma^{}_\mathrm{L}$ are the measured source peak flux and local background noise level in the image. For a fixed upper limit on the calibration time interval, an increase in $\alpha$ results in a decrease in the number of peeling sources. For $n^{}_t < 1$, phase corrections are determined on the visibility time grid. For $n^{}_t > 1$, a spline is used to resample the phase corrections per antenna in time onto the visibility time grid. 

Apart from SNR issues, the number of sources that can be peeled is fundamentally limited by the available number of independent visibility measurements. When peeling $N^{}_{s}$ sources, self-calibration fits $N^{}_{s} ( N^{}_{a} - 1 )$ phase solutions per calibration time interval to the visibility data, where $N^{}_a$ is the number of antennas. For self-calibration to converge to an unique combination of phase solutions and source model, this number needs to be much smaller than the number of independent visibility measurements. The maximum of visibilities measurements that is available in one calibration time interval is given by $N^{}_{c} \langle n^{}_{t} \rangle N^{}_{a} ( N^{}_{a} - 1 ) / 2$, with $N^{}_{c}$ the number of frequency channels and $\langle n^{}_t \rangle$ the average number of visibility integration times in a calibration interval. In the ideal case, when we assume that each visibility is an independent measurement, the determination of antenna-based phase corrections for all peeling sources is well constrained if
\begin{equation}
N^{}_{s} \ll \frac{ N^{}_{a} N^{}_{c} \langle n^{}_{t} \rangle }{ 2 }.
\label{eq:peel_number_limit}
\end{equation}
The applications presented in this article do satisfy this minimal condition (see Section~\ref{sec:apps}). 

Equation~\ref{eq:peel_number_limit} is equivalent to stating that the number of degrees-of-freedom (DoF; the difference between the number of independent measurements and the number of model parameters) should remain a large positive number. Correlation between visibilities over frequency and time may reduce the number of independent measurements drastically, thereby also reducing the number of DoFs. The exact number of DoFs for any data set is hard to quantify. When this number becomes too low, the data is `over-fitted' (e.g., Bhatnagar et al. \cite{bib:bhatnagar2008}), which could result in an artificial reduction of both the image background noise level and source flux that is not represented in the self-calibration model (Wieringa \cite{bib:wieringa1992}). Although we have found no evidence of this effect occuring in the applications presented in this article, the SPAM user should be cautious not to peel too many sources. In case of a high number of available peeling sources, one can choose a subset with a sufficiently dense spatial distribution over the FoV (e.g., one source per isoplanatic patch; see Section~\ref{sec:imaging}).

\subsection{Ionospheric Phase Screen Model}
\label{sec:iono_model}

The phase corrections that are obtained by peeling several bright sources in the FoV (Section~\ref{sec:peeling}) are only valid for ionospheric calibration in a limited patch of sky around each source. To correct for ionospheric phase errors over the full FoV during wide-field imaging and deconvolution, SPAM requires a model that predicts the phase correction per antenna for arbitrary viewing directions.

\begin{figure}
\begin{center}
\resizebox{\hsize}{!}{\includegraphics[angle=0]{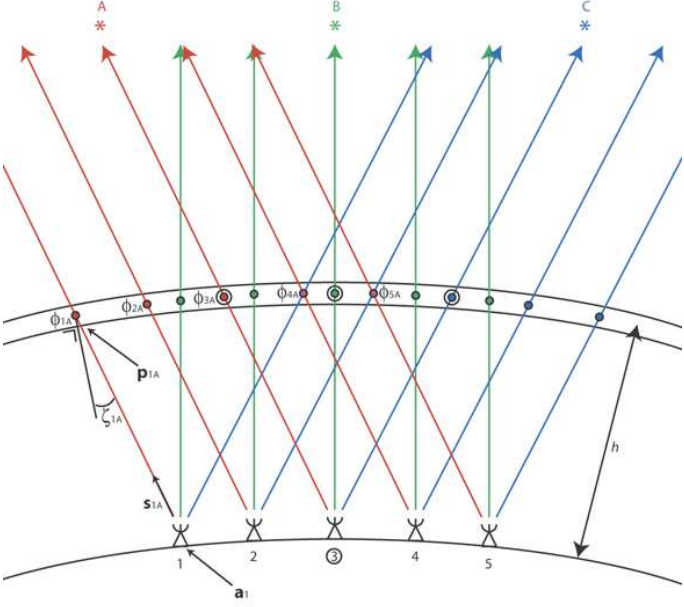}}
\caption{Schematic overview of the SPAM thin ionospheric phase screen model geometry. For clarity, only two spatial dimensions and one calibration time interval are considered. In this overview, five ground-based array antennas (labelled 1 to 5) observe three calibrator sources (colored red/green/blue and labelled A to C) within the FoV. The (colored) LoSs from the array towards the sources run parallel for each source and pierce the phase screen at fixed height $h$ (colored circles). The LoS from antenna $i$ at Earth location $\vec{a}^{}_i$ towards a peeling source $k$ at local sky position $\hat{\mathbf{s}}^{}_{ik}$ intersects the phase screen at a single \textit{pierce point} $\vec{p}^{}_{ik}$ under a zenith angle $\zeta^{}_{ik}$. For a single LoS from antenna~1 towards source~A, we have indicated how the pierce point position $\vec{p}^{}_{ik} = \mathbf{p}^{}_\mathrm{1A}$ and zenith angle $\zeta^{}_{ik} = \zeta^{}_\mathrm{1A}$ relate to the antenna position $\vec{a}^{}_{i} = \mathbf{a}^{}_\mathrm{1}$ and the local sky position $\hat{s}^{}_{ik} = \mathbf{s}^{}_\mathrm{1A}$ of the source. For some LoSs the pierce points may overlap (or nearly overlap), as is the case for 1C~\&~4A and 2C~\&~5A in our example. The total (integrated) phase rotation along any LoS through the ionosphere is modeled by an instantaneous phase rotation $\phi^\mathrm{ion}_{ik}$ at the phase screen height. For example, radio waves traveling along LoSs from source~A towards antennas 1 to 5 experience an instantaneous phase rotation $\phi^\mathrm{ion}_{ik} = \phi^{}_\mathrm{1A}$ to $\phi^{}_\mathrm{5A}$, respectively, while passing the screen at their related pierce points $\vec{p}^{}_{ik} = \mathbf{p}^{}_\mathrm{1A}$ to $\mathbf{p}^{}_\mathrm{5A}$, respectively. Peeling the three calibrator sources yields measurements of the ionospheric phases $\phi^\mathrm{ion}_{ik}$, relative to a common reference antenna (in this example antenna 3; encircled).}
\label{fig:pierce_points}
\end{center}
\end{figure}

SPAM constructs a quasi-physical phase screen model that attempts to accurately reproduce and interpolate the measured ionospheric phase rotations (or more accurately: the peeling phase corrections). The phase screen is determined independently for each visibility time stamp, therefore we drop the $n$-subscript in the description below. Figure~\ref{fig:pierce_points} is a schematic overview of the geometry of ionospheric phase modeling in SPAM. The ionosphere is represented by a curved phase screen at a fixed height $h$ above the Earth's surface, compliant to the WGS84 standard (NIMA \cite{bib:wgs1984}). The total phase rotation experienced by a ray of radio emission traveling along a LoS through the ionosphere is represented by an instantaneous phase rotation $\phi^\mathrm{ion}_{}(\vec{p},\zeta)$ on passage through the phase screen that is a function of pierce point position $\vec{p}$ and zenith angle $\zeta$. For a thin layer ($\Delta h \ll S$; see Section~\ref{sec:calib}), the dependence of $\phi^\mathrm{ion}_{}$ on $\zeta$ can be represented by a simple airmass function, so that
\begin{equation}
\phi^\mathrm{ion}_{}(\vec{p},\zeta) = \frac{ \phi^\mathrm{ion}_{}(\vec{p}) }{ \cos(\zeta) }.
\label{eq:airmass}
\end{equation}

SPAM uses an angular local longitude/latitude coordinate system to specify $\vec{p}$, relative to the central pierce point from array center to field center. For the applications presented in this article, the angular distances between pierce points over the relevant ionospheric domain are all $< 5$~degrees, which effectively makes the pierce point vector $\vec{p}$ a 2-dimensional cartesian vector.

\begin{figure*}
\begin{center}
\resizebox{\hsize}{!}{\includegraphics[angle=0]{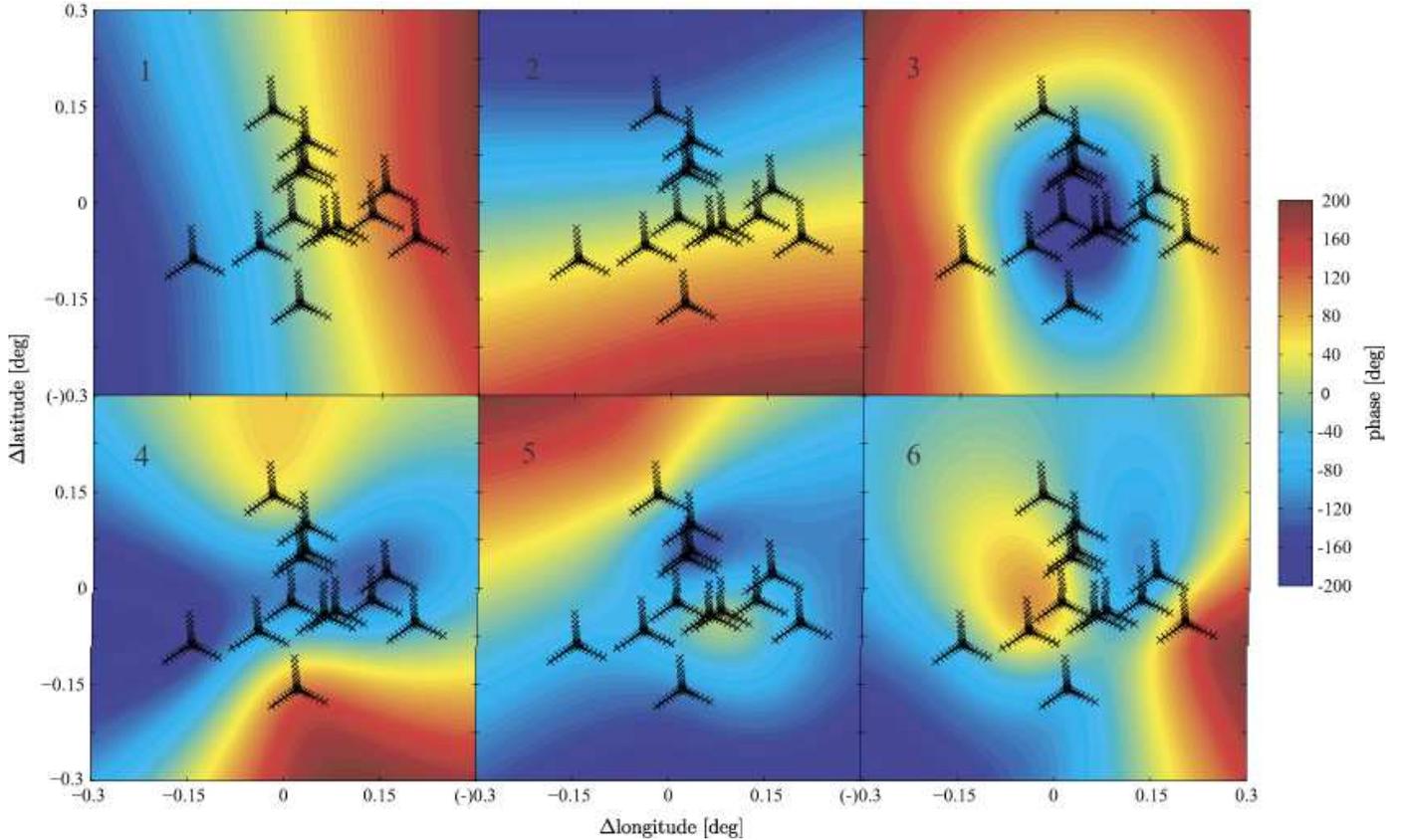}}
\caption{Plots of the interpolations of the first six KL base vectors, derived for an artificial but realistic configuration of ionospheric pierce points. In this example, the pierce points (black crosses) are calculated for a single time instance during a 74~MHz VLA-B observation with 13~available calibrator sources in the $\sim 10$~degree FoV, adopting a phase screen height $h = 200$~km and a structure function power-law slope $\gamma = 5/3$. The horizontal and vertical axes represent angular distances in East-West and North-South directions, respectively, as seen from the center of the Earth, relative to the phase screen's pierce point along the line-of-sight from array center to pointing center, with East- and Northward offsets being positive. At this height, a 0.1~degree angular offset represents a physical horizontal offset of $\sim 11.5$~km. The direction-dependent phase for each interpolated KL base vector is color-coded and scaled to an arbitrary amplitude range.}
\label{fig:kl1to6}
\end{center}
\end{figure*}

The 2-dimensional phase screen $\phi^\mathrm{ion}_{}(\vec{p})$ is defined on a set of KL base vectors, generated from the instantaneous pierce point configuration $\{\vec{p}^{}_{ik}\}$ and an assumed power-law shape for the phase structure function (Section~\ref{sec:ionosphere}). The KL base vector generation and interpolation is based on the work by van der Tol \& van der Veen (\cite{bib:vdtolvdveen2007}) and is described in detail in Appendix~\ref{app:kl}. The phase screen model requires one free parameter per KL base vector. The initial complete set of KL base vectors is arbitrarily reduced in order by selecting a subset based on statistical relevance (principle component analysis). This reduces the effect of noise in the peeling solutions on the model accuracy and simultaneously limits the number of model parameters. However, the subset should still be large enough to accurately reproduce the peeling phase corrections. Per visibility time stamp, the KL base vectors are stored for later use during imaging (for this purpose, we mis-use the AIPS OB table). As an example, the first six interpolated KL base vectors for a single configuration of ionospheric pierce points are plotted in Figure~\ref{fig:kl1to6}.

The peeling phase corrections $\phi^\mathrm{cal}_{ik}$ are interpreted to be relative measurements of the absolute ionospheric phase screen model $\phi^\mathrm{ion}_{}(\vec{p},\zeta)$ which may be determined up to a constant. The model parameters are determined by minimizing the differences between the observed and the model phases using the LM non-linear least-squares solver, for which a $\chi^{2}_{}$ sum needs to be defined. From Equation~\ref{eq:peel_phase} it follows that
\begin{equation}
\phi^\mathrm{cal}_{ik} = \phi^\mathrm{ion}_{}(\vec{p}^{}_{ik},\zeta^{}_{ik}) - \phi^\mathrm{ion}_{}(\vec{p}^{}_{rk},\zeta^{}_{rk}) - \phi^\mathrm{ambig}_{ik}.
\label{eq:ion_model_ir}
\end{equation}
Consequently, the phase correction in the direction of source $k$ for a baseline consisting of antennas $i$ and $j$ is
\begin{eqnarray}
\phi^\mathrm{cal}_{ik} - \phi^\mathrm{cal}_{jk} & = & \big[ \phi^\mathrm{ion}_{}(\vec{p}^{}_{ik},\zeta^{}_{ik}) - \phi^\mathrm{ion}_{}(\vec{p}^{}_{jk},\zeta^{}_{jk}) \big] - \nonumber \\
& & \quad \quad \big[ \phi^\mathrm{ambig}_{ik} - \phi^\mathrm{ambig}_{jk} \big].
\label{eq:ion_model_ij}
\end{eqnarray}
The $\chi^{2}_{}$ sum is defined as:
\begin{eqnarray}
\chi^{2}_{} & = & \displaystyle \sum^{}_{k} \sum^{}_{i} \sum^{}_{j>i} \bigg[ \Big( \big[ \phi^{cal}_{ik} - \phi^{cal}_{jk} \big] \, - \nonumber \\
& & \quad \big[ \phi^\mathrm{ion}_{}(\vec{p}^{}_{ik},\zeta^{}_{ik}) - \phi^\mathrm{ion}_{}(\vec{p}^{}_{jk},\zeta^{}_{jk}) \big] \Big) \bmod{ 2 \pi } \bigg]^2.
\label{eq:chi_squared}
\end{eqnarray}
This definition has several properties: (i) By remapping the $\chi^{2}_{}$ terms into the $[0,2\pi)$ domain, the phase ambiguity terms do not have to be fitted explicitly, (ii) the $\chi^{2}_{}$ terms of all calibrator sources are weighted equally, so the model is not biased towards the brightest source (as is the case for self-calibration), and (iii) using $\chi^{2}_{}$ terms from all possible antenna pairs prevents a bias towards the reference antenna. 

Using Equation~\ref{eq:chi_squared}, the LM solver yields a set of model parameters per visibility time stamp. These are stored for later use during imaging (AIPS NI table). The square root of the average of the $\chi^{2}_{}$ terms equals the average RMS phase residual between peeling and model phases. Time intervals that have a bad fit are identified and removed by means of an upper limit ($+ 2.5\,\sigma$ rejection) on the distribution of RMS phase residuals over time. 

Convergence of the LM solver is troubled by $2 \pi$ phase ambiguities, because these introduce local minima in $\chi^{2}_{}$ space. A good initial guess of the model parameters greatly helps to overcome this problem. To this purpose, SPAM estimates the global phase gradient over all the pierce points directly from the phase corrections $\phi^\mathrm{cal}_{ik}$ and projects it onto the KL base vectors before invoking the LM solver.

\begin{figure}
\begin{center}
\resizebox{\hsize}{!}{\includegraphics[angle=0]{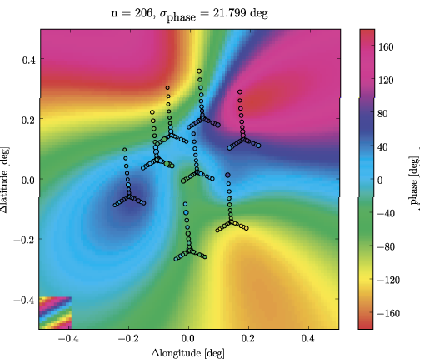}}
\caption{Example of an ionospheric phase screen model fit. The color map represents an ionospheric phase screen at 200~km height that was fitted to the peeling phase solutions of 8 calibrator sources at time-interval $n = 206$ of 10~seconds during a VLSS observing run of the 74 MHz VLA in BnA-configuration (see Section~\ref{sec:apps}, the J1300-208 data set). The plot layout is similar to Figure~\ref{fig:kl1to6}. The overall phase gradient (depicted in the bottom-left corner) was removed to make the higher order terms more clearly visible. The collection of pierce points from all array antennas to all peeling sources are depicted as small circles., The color in the circle represents the measured peeling phase (the reference antenna VLA N36 was set to match the phase screen value). The size of the circle scales with the magnitude of the estimated phase residual after model correction. The overall RMS phase residual $\sigma^{}_\mathrm{phase} = 21.799$~degrees (averaged over all pierce points) was one of the better fitting results during this particular observing run.}
\label{fig:model_fit_example}
\end{center}
\end{figure}

\begin{figure}
\begin{center}
\resizebox{\hsize}{!}{\includegraphics[angle=0]{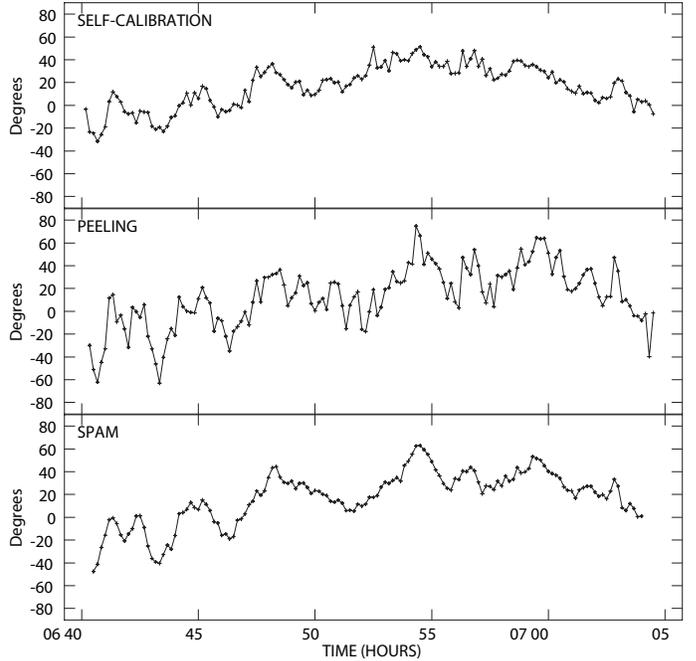}}
\caption{Example of phase corrections from different steps in the ionospheric calibration process, resulting from processing a VLSS data set with SPAM (see Section~\ref{sec:apps}, the real J0900+398 data set). The antenna under consideration is VLA E28, with W20 being the reference antenna (an 5.7~km east-west baseline). The plots represent 25~minutes of observing time, using a 10~second time resolution. \textit{Top}: Antenna-based phase corrections resulting from self-calibration on the whole FoV. \textit{Middle}: Phase corrections resulting from peeling the brightest (30~Jy) source. \textit{Bottom}: Corrections resulting from ionospheric phase modeling in the direction of the (same) brightest source.}
\label{fig:example_sc_peel_spam_phases}
\end{center}
\end{figure}

Figure~\ref{fig:model_fit_example} shows an example of an ionospheric phase screen that was constructed as described above. The pierce point layout consists of multiple projections of the array onto the phase screen. The low density of calibrators causes a minimal overlap between array projections. Figure~\ref{fig:example_sc_peel_spam_phases} shows a comparison between time-sequences of phase corrections from self-calibration, peeling and model fitting. Because the self-calibration corrections are a flux-weighted average for the full FoV, they are biased towards the brightest source. They look somewhat similar to the peeling solutions of the brightest source, but the latter contains additional fluctuations that vary on a relatively short timescale. The model phases appear similar to the peeling phases, but vary more smoothly. Their values fall somewhere in between the self-calibration phases and the peeling phases. The difference between the peeling phases and model phases are mainly caused by the constraints on the spatial variability of the phase screen model.

\subsection{Imaging}
\label{sec:imaging}

With an ionospheric phase screen model available for a given visibility data set, antenna-based phase corrections for any direction in the wide FoV can be calculated (Equation~\ref{eq:ion_model_ir}). Because each visibility consist of contributions from visible sources in different viewing directions, there is no simple operation that removes the ionospheric phase rotations from a visibility data set prior to imaging. Instead, SPAM requires an algorithm that calculates and applies the appropriate model phase corrections during imaging and deconvolving for different parts of the FoV.

SPAM works under the assumption that there exists a fixed angular \textit{isoplanatic patch} size on the sky, with a projected size at ionospheric height smaller than the scale size of ionospheric phase fluctuations, over which variations in ionospheric phase rotation are negligible. Each isoplanatic patch requires at least one phase correction per antenna per visibility time interval. For the VLA at 74 MHz, the isoplanatic patch size is estimated to be 2--4 degrees (Cotton \& Condon \cite{bib:cottoncondon2002}).

The facet-based polyhedron method for wide-field imaging (see Section~\ref{sec:init_sky_calib}) allows for a relatively simple implementation of ionospheric phase correction (Schwab \cite{bib:schwab1984}). By choosing a facet size smaller than the isoplanatic patch size, a set of model phase corrections calculated for the center of a facet are assumed to be accurate for the whole facet area. Ionospheric phase model corrections are calculated and stored (AIPS SN tables) for each facet center in the FoV prior to imaging and deconvolution. For the additional facets centered on bright sources (see Section~\ref{sec:init_sky_calib}), model phase corrections are optionally replaced by peeling phase corrections to allow for optimized calibration towards these sources.

The SPAM imaging and deconvolution procedure is similar to the procedure used for the field-based calibration method by Cotton et al. (\cite{bib:cotton2004}), which differs from the standard Cotton-Schwab algorithm by the temporary application of the facet-based phase corrections (AIPS tasks SPLIT and CLINV/SPLIT) to the visibility data for the duration of major CLEAN cycles on individual facets (AIPS tasks IMAGR and UVSUB). After deconvolution, facets are combined to form a single image of the full FoV (AIPS task FLATN). Because antenna-based phase corrections change very little between adjacent facets, the complete set of partly overlapping facet images combine into a continuous image of the FoV.

\section{Applications}
\label{sec:apps}

To demonstrate the capabilities of SPAM, we have defined three test cases based on observations with the VLA at 74~MHz (Kassim et al. \cite{bib:kassim2007}). In each test case, SPAM is used for ionospheric phase calibration and imaging of a VLSS visibility data set (Cohen et al. \cite{bib:cohen2007}), following the steps described in Section~\ref{sec:method}. In the first test case, SPAM was applied to simulated data to validate basic functionality in a controlled environment. In the next two test cases, SPAM was applied to visibility data from real observations under varying ionospheric conditions. We compare SPAM performance against self-calibration (SC) and field-based calibration (FBC) by analyzing the resulting images. The setup and results of these test cases are described in detail in the following sections.

\subsection{Data Selection, Preparation and Processing}
\label{sec:sel_prep}

\ctable[botcap,center,star,
caption = {Overview of processing parameters for the three data sets that are handled with SPAM as defined in the test cases.},
label = tab:processing
]{l c c c}{
\tnote[a]{The pixel size for all field-based calibration images is 20\arcsec.}
\tnote[b]{Adding more cycles did not significantly improve the image quality.}
\tnote[c]{Arbitrarily limited to mimic a more realistic scenario.}
\tnote[d]{Increased to improve match with FBC phase screen.}
\tnote[e]{In this case, 15 terms proved to be insufficient.}
}{\FL
Field name & VLSS J0900+398 (simulated) & VLSS J0900+398 (real) & VLSS J1300-208 (real) \ML
Pixel size\tmark[a] & 18.9\arcsec & 18.9\arcsec & 11.1\arcsec \NN
Number of facets & 347 & 243 & 576 \NN
Facet separation & $1.18\degs$ & $1.18\degs$ & $0.62\degs$ \NN
SPAM calibration cycles\tmark[b] & 1 & 1 & 3 \NN
Peeling sources & 10\tmark[c] & 11 & 9 \NN
KL model height & 1000 km\tmark[d] & 200 km & 200 km \NN
Fitted KL model terms & 15 & 15 & 20\tmark[e] \NN
Rejected time intervals & 0 / 464 & 25 / 464 & 86 / 484 \NN
Model fit phase RMS & $3.0\degs \pm 0.8\degs$ & $21.3\degs \pm 2.4\degs$ & $23.2\degs \pm 3.2\degs$ \NN
Peeling corrections applied directly & no & yes & yes \LL
}

In this Section we describe how the visibility data sets for the three test cases were selected/constructed. Furthermore, we present details on how these data sets were processed by SPAM into calibrated images of the FoV.

Two VLSS observations, at pointing centers J0900+398 and J1300-208, respectively, have been picked from more than 500 available VLSS observations on the following criteria: (i) both fields contain a relatively large number of bright sources that can serve as calibrators, and (ii) the ionospheric conditions during the observations appear to be relatively good (J0900+398) and relatively bad (J1300-208). The presence of more than 5~bright sources of at least 5~Jy compensates for the relatively poor efficiency of the VLA 74~MHz receiving system (Kassim et al.~\cite{bib:kassim2007}). The ionospheric conditions were derived from the apparent smearing of point sources in the images, due to residual phase errors after applying FBC. From experience, we adopted the qualification `good' when the mean width of apparent point sources was at most 5\arcsec{} larger than the intrinsic 80\arcsec{} resolution, while for `bad' conditions the mean point source width was larger by at least 15\arcsec. In terms of Strehl ratio $R$ (Equation~\ref{eq:strehl_ratio}), 'good' and 'bad' conditions correspond with $R > 0.996$ and $R < 0.966$, respectively. Additionally, candidate fields were visually inspected for evidence of residual phase errors by the presence or absence of image artefacts near bright sources, which lead to the final selection of the two fields mentioned above.

The difference in observed ionospheric conditions between the two real data sets may be the result of the difference in array size and elevation of the target field. From the VLA site at +34~degrees declination, the J0900+398 field was observed in B-configuration (up to 11~km baselines) at relatively high elevation, while the J1300-208 field was observed in BnA-configuration (up to 23~km baselines) at relatively low elevation. For the J1300-208 observation, the array observed through the ionosphere at larger separations and along longer path lengths than for the J0900+398 observation, which is expected to result in both larger and less coherent phase errors over the array.

Because both real data sets have been previously calibrated and imaged with FBC, the data sets were already partly reduced at the start of SPAM processing. Instrumental calibration was applied (including instrumental phase calibration, similar to Section~\ref{sec:instr_calib}), most RFI-contaminated data was flagged and the spectral resolution was reduced (see Cohen et al. \cite{bib:cohen2007} for details), but no FBC has been applied yet. For the simulated data set, which is based on the real J0900+398 observations, the measured visibilities were replaced by noiseless model visibilities of an idealized sky, consisting of 91~bright point sources with peak fluxes (larger than 1~Jy) and positions as measured in the J0900+398 FBC image. For each point source, the corresponding model visibility phases were corrupted using the direction-dependent ionospheric phase model that was obtained with FBC to correct the real J0900+398 data.

FBC images of the two real data sets were available in the VLSS archive. For the simulated J0900+398 data set, an `undisturbed' image was made before applying the ionospheric phase corruptions. All three VLSS data sets have been processed with SPAM, yielding both an SC image and an ionosphere-corrected SPAM image. Relevant details on the processing can be found in Table~\ref{tab:processing}. For SC and SPAM imaging, we adopted most of the imaging-specific settings from FBC (like uniform weighting). Noticeable differences are the use of CLEAN boxes, a smaller pixel size and a different facet configuration.

By choosing a minimum SNR per time interval of 15 and a maximum peeling time interval of 4~minutes (see Equation~\ref{eq:peel_time_limit}), SPAM was able to peel $\sim 10$ sources in each of the real data sets. Lowering the SNR resulted in a much larger scatter in the peeling phases over time, or prevented peeling to converge at all. The peeling time upper limit was chosen to roughly match the spatial density of calibrator sources used in FBC. Determining phase corrections on a 4~minute time scale could result in undersampling the time evolution of ionospheric phase errors. Note that this only applies to the faintest of the calibrator sources. The limitations on spatial and temporal sampling of the ionosphere are dictated by the given sensitivity of the VLA.

Because of the high SNR, all 91~sources in the simulated J0900+398 data set qualified for peeling at the highest time resolution of 10~seconds. To mimic a more realistic scenario for further SPAM processing, the number of calibrators was arbitrarily limited to 10. Generally, for all data sets, the images of peeling sources showed larger peak fluxes and less background structure than their counterparts in the SC image, although the contrast became less apparent for weaker and extended (mostly doubles) peeling sources.

As stated in Section~\ref{sec:peeling}, the number of peeling sources is fundamentally limited by the requirement for a large positive number of degrees-of-freedom in the available visibility data. The minimal requirement is given in Equation~\ref{eq:peel_number_limit}. Typically, for the VLSS data sets, there were 25~active antennas, 12~frequency channels and 6~visibility intervals (of 10~seconds) in an average peeling interval of 1 minute. In our test cases, we typically peel 10~sources, which is much less than $25 \times 12 \times 6 / 2 = 900$, thereby satisfying the minimal requirement.

Due to the uncertainty in their optimal values, it is left to the SPAM user to specify the phase screen model order (the number of KL base vectors), the height $h$ of the phase screen and the power-law exponent $\gamma$ of the phase structure function. For the applications presented here, we used $h = 200$~km and $\gamma = 5 / 3$, which is compliant to the measured values given in Section~\ref{sec:ionosphere} given the uncertainty in these values. For the simulated data set, we chose instead $h = 1000$~km to better match the corrupting FBC ionospheric phase model that is attached to the sky plane at infinite height. These values gave satisfactory results for the test applications presented here, but can be further optimized. The optimal model order was found to lie in the range of 15--20 terms, which is 1.5--2 times the number of available peeling sources. Increasing or decreasing the model order caused the model fit to be less accurate or more problematic in terms of convergence.

For both the simulated and real J0900+398 data sets, no improvement in background noise was observed by adding a second calibration cycle after the first. This indicates fast convergence of the SPAM calibration method for quiet ionospheric conditions, where the initial self-calibration is already close to the best achievable calibration of SPAM. For the real J1300-208 data set, adding up to third calibration cycle did improve over the previous cycles.

\subsection{Phase Calibration Accuracy}
\label{sec:phase}

\begin{figure}
\begin{center}
\resizebox{\hsize}{!}{\includegraphics[angle=0]{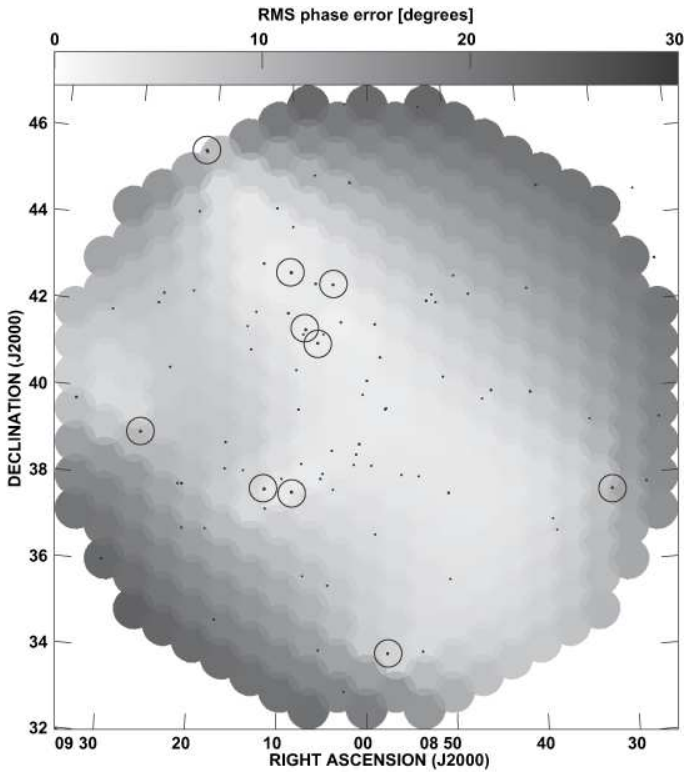}}
\caption{The grayscale map represents the residual phase RMS between the distorting and correcting ionospheric phase models across the primary beam area, averaged over baselines and time. The phase RMS was calculated for a hexagonal grid of viewing directions across the FoV. Each viewing direction is depicted by a small circular area. Overplotted is a contour map of the point sources as seen in the SPAM image (which extends slightly beyond the grid of circles). The 10~peeling sources are marked by circles. The correspondence between the models is largest near the calibrator sources and over a large part of the inner primary beam. The discrepancy is largest near the South-East and North-West borders, away from the calibrators.}
\label{fig:model_sources_errors}
\end{center}
\end{figure}

For the simulated J0900+398 data set, the absolute accuracy of ionospheric calibration can be determined by a direct comparison between the corrupting FBC phase screen and the correcting SPAM phase screen. To this purpose, phase corruptions and corrections were calculated from the models for a hexagonal grid of 342~viewing directions within the FoV. Per viewing direction, the RMS phase error was calculated by differencing of the phases from both models and averaging over all time stamps and baselines. The result is depicted in Figure~\ref{fig:model_sources_errors}.

For areas near the calibrators and in the center of the field in general, there is a relatively good match between the input and output model, with typical RMS phase errors $\lesssim 5$~degrees. The absence of calibrator sources south-west of the field center still results in relatively accurate predictions by the SPAM model. In the direction of peeling sources, the measured RMS phase error can be split into a contribution from inaccuracies in the peeling process and a contribution from imperfect model fitting. The latter is approximately 3~degrees (Table~\ref{tab:processing}), therefore the RMS phase error introduced by peeling is $\lesssim 4$~degrees. Considering the model setup, the only possible source of error is contamination from other sources while peeling (which appears to happen despite the initial subtraction of the SC model).

Overall, the change in model base from the corrupting FBC model (5~Zernike polynomials) to the correcting SPAM model (15~KL vectors) has a constant accuracy over large parts of the FoV. Towards some parts of the edge of the field the phase errors are substantially larger, up to 20--25~degrees at worst. This agrees with the different asymptotic behaviour towards large radii of the Zernike model (diverge to infinity) and the KL model (converge to zero) in the absence of calibrators. The presence of calibrator sources near the edge (like the one on the North-East edge of the field) leads to a better local match between corrupting ionosphere and correcting model. 

For the real observations, in the absence of external sources of information (e.g., GPS measurements), it is not possible to derive the absolute accuracy of ionospheric calibration from the observations themselves. Instead, the residual RMS phase error of the model fit to the peeling phases is used as an relative indicator for calibration accuracy over time. For both the real J0900+398 and J1300-208 data sets, the residual RMS phase error of $\sim 22$~degrees is much larger than for the simulated data. This already excludes rejected time stamps with exceptionally large RMS values. By inspecting model fits on individual time stamps, we found that there are often a few pierce point phases that deviate significantly more from the fitted model than most neighbouring points. These errors do not appear to be antenna-based instrumental errors, because peeling solutions for the same antenna towards other calibrator sources do not deviate in the same manner. Typically, these deviating points persist for a few time stamps before disappearing. The ionosphere may be responsible for these very small scale deviations. Another possibility is that the peeling solutions are (sometimes) noisy due to limitations in source SNR.

\subsection{Background Noise}
\label{sec:noise}

\ctable[botcap,center,star,
caption={Overview of results from calibrating and imaging three test case data sets with no ionosphere (Undisturbed), self-calibration (SC), field-based calibration (FBC) and SPAM.},
label=tab:results
]{l c c c}{
}{\FL
Field name & VLSS J0900+398 (simulated) & VLSS J0900+398 (real) & VLSS J1300-208 (real) \ML
\multicolumn{4}{l}{Mean background noise $\sigma$ [\mjybeam]:} \NN
Undisturbed &  3.0 & -- &  -- \NN
SC          & 10.2 & 71 &  92 \NN
FBC         &   -- & 87 & 118 \NN
SPAM        &  6.7 & 67 &  68 \ML
\multicolumn{4}{l}{Number of sources with a peak flux larger than $5 \sigma$:} \NN
Undisturbed & 91 &  -- &  -- \NN
SC          & 91 & 393 & 374 \NN
FBC         & -- & 310 & 285 \NN
SPAM        & 91 & 372 & 392 \ML
\multicolumn{4}{l}{$5 \sigma$ source fraction with an NVSS counterpart within 80\arcsec:} \NN
Undisturbed & 1. &   -- &   -- \NN
SC          & 1. & 0.83 & 0.60 \NN
FBC         & -- & 0.86 & 0.74 \NN
SPAM        & 1. & 0.97 & 0.97 \LL
}

In this and the next sections, we revert to analyzing image properties for an indirect, relative comparison between the different calibration techniques. In the presence of residual phase errors, part of the image background noise level consists of residual sidelobes after CLEANing. The local sidelobe noise increases with both the RMS phase error and the local source flux. When measured over a large image area, the mean sidelobe noise depends mainly on RMS phase error. For all relevant output images, the mean image noise $\sigma$ was determined by fitting a Gaussian to the histogram of image pixel values from the inner quarter radius of the FoV (AIPS task IMEAN). Note that these images have not been corrected for primary beam attenuation. The results are given in Table~\ref{tab:results}.

Because no noise was added to the simulated J0900+398 data set, the resulting image noise of $3.0$~\mjybeam in the undisturbed image is caused by incomplete UV coverage and inaccuracies in the imaging process (see Section~\ref{sec:init_sky_calib}), limiting the dynamic range to $\sim 10^{4}_{}$. The local noise is highest near the sources, but significantly less near the brightest 10~sources with dedicated facets centered on their peak position. The SC and SPAM images from this data set were created using the same facet configuration. The SC image noise of $10.2$~\mjybeam is 3.4 times as high as the undisturbed image noise, therefore dominated by phase error induced sidelobe noise. The SPAM image noise of $6.7$~\mjybeam is a significant improvement over the SC image, but still 2.2~times as high as in the undisturbed image. The local noise in the SC and SPAM images has increased most apparently near bright sources as compared to the undisturbed image, which confirms the presence of residual phase errors after calibration.

\begin{figure*}
\begin{center}
\resizebox{\hsize}{!}{\includegraphics[angle=0]{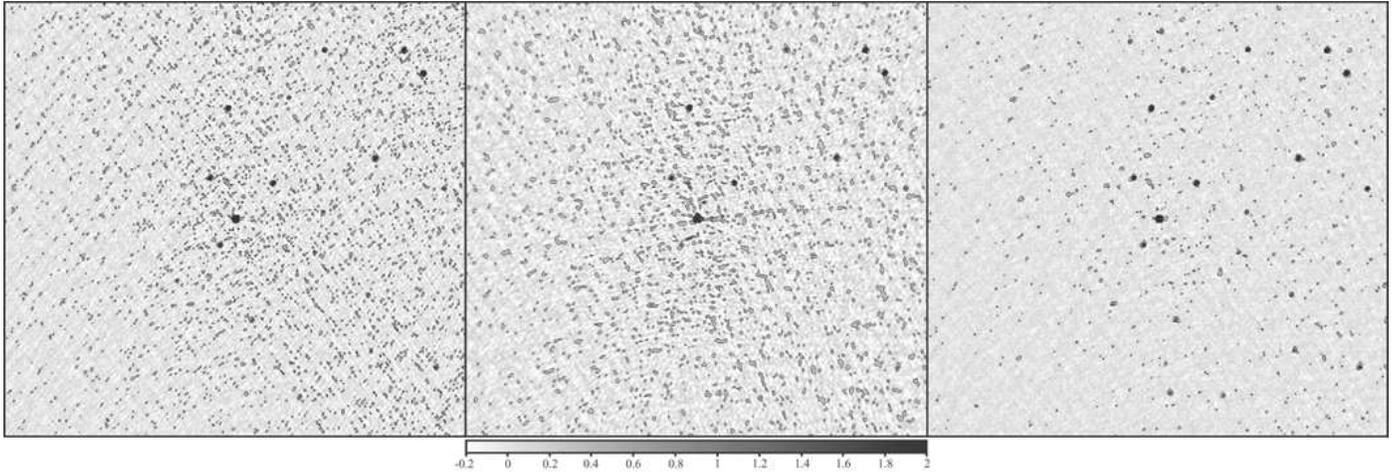}}
\caption{Greyscale plots of a $3.5 \times 3.5$ square degree area in the VLSS~J1300-208 field centered on the bright (40~Jy) point source 3C~283. All three images have contours (black lines) overplotted at [0.15, 0.48, 0.83, 1.16, 1.50]~Jy.  \textit{Left}: Image after self-calibration, \textit{middle}:  image after field-based calibration, and \textit{right}: image after SPAM calibration.}
\label{fig:combined_bright_source}
\end{center}
\end{figure*}

For the real J0900+398 data set, both the SC and SPAM images have an image noise of $\sim 70$~\mjybeam. The SPAM image noise is slightly lower than SC. The local noise in the SC image is higher near bright sources. This is not the case in the SPAM image, which must be a direct result of an improved calibration accuracy near these sources. The FBC image noise for this data set is $\sim 20$~percent higher, a combination of a higher average noise over the FoV and higher local noise near bright sources.

For the real J1300-208 data set, the SPAM image has the same image noise as for the real J0900+398 data set, with no apparent increase near bright sources. At the same time, the noise levels in the SC and FBC images have increased with 30 and 35~percent, respectively. The noise in the SC image is highest near the bright sources. The FBC noise is highest near the brightest source and remains high in the rest of the image. The significant increase of the average FBC noise level indicates a dependence on ionospheric conditions, and therefore on calibration accuracy. The SPAM image noise appears to have little or no dependence on varying ionospheric conditions (Figure~\ref{fig:combined_bright_source}).

\subsection{Source Properties}
\label{sec:src_prop}

The presence of residual phase errors changes the apparent distribution of flux of a source (see Section~\ref{sec:img_effects}). In the time-averaged image, sources may appear offset from their intrinsic position, may suffer from smearing or deformation, and sidelobes may be misidentified as sources. Comparing the properties of the same sources in differently calibrated images allows for a relative comparison of the performance of the different calibration techniques.

To allow for comparison of source properties, we applied the source extraction tool BDSM (Mohan et al. \cite{bib:mohan2008}) on all relevant images. BDSM performed a multiple 2-dimensional Gaussian fit on islands of adjacent pixels with amplitudes above a specified threshold based on the \textit{local} image noise $\sigma^{}_\mathrm{L}$ in the image. Multiple overlapping Gaussians were grouped together into single sources. We applied BDSM to all images, using the default extraction criteria, except for the following: a source detection requires at least 4~adjacent pixel values above $2.5\,\sigma^{}_\mathrm{L}$, with at least one pixel value above $4\,\sigma^{}_\mathrm{L}$.

\begin{figure}
\begin{center}
\resizebox{\hsize}{!}{\includegraphics[angle=0]{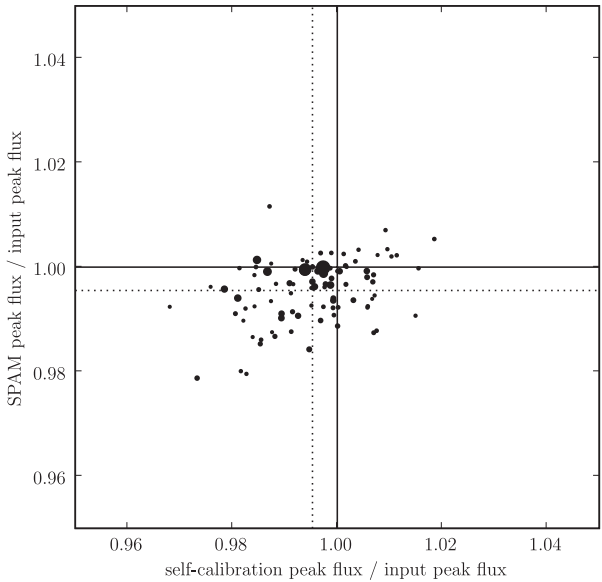}}
\caption{Peak flux ratios of point sources in the simulated J0900+398 field. Peak fluxes were measured in the self-calibration image and the SPAM image, corrected for a small CLEAN bias and divided by the input model peak fluxes. The size of each dot scales with the input model peak flux, ranging from 1.02 to 26.7~Jy. Ideally (without phase errors), the peak flux ratios would be scattered around \textit{one} (solid lines) due to image noise dependent errors in the peak flux determination. Instead, the peak flux ratio distributions along the x- and y-axis are centered around 0.995 and 0.996, respectively (dotted lines), which is a direct result of the residual phase errors. The smaller and larger scatter in distribution of SPAM- and self-calibration peak flux ratios is consistent with peak flux determination inaccuracies due image background noise levels.}
\label{fig:model_spam_sc_peak_peak}
\end{center}
\end{figure}

\subsubsection{Source Counts}
\label{sec:src_counts}

Due to the non-Gaussian character of the phase-induced sidelobe noise, the source catalogs will contain spurious detections. To suppress these, we removed sources with a peak flux smaller than $5\,\sigma$ from the catalogs. The remaining number of catalog entries are listed in Table~\ref{tab:results}. Additionally, each catalog was cross-associated against the NVSS catalog, which has a slightly higher resolution (45\arcsec). For an average spectral index of $- 0.8$, the NVSS detection limit is at least 10 times lower than for the VLSS. At the risk of missing an incidental ultra-steep spectrum source, we determined the source fraction that has an NVSS counterpart within an 80\arcsec{} radius (one VLSS beamsize), which are also listed in Table~\ref{tab:results}.

For the simulated J0900+398 data set, all 91~input sources are detected and matched against NVSS counterparts, regardless of the calibration method. Due to the low noise levels and the lower limit of 1~Jy on the input source catalog, all sources are effectively $\gtrsim 100\,\sigma$ detections. None of the sources had more than one Gaussian fitted to it, despite the freedom to do so.

For the real J0900+398 data set, the higher $\sigma$ in the FBC image is reflected in a smaller number of source detections as compared to SC and SPAM. SC detects slightly more sources than SPAM, despite the slightly higher $\sigma$. However, there is a very large fraction of sources in the SPAM catalog that has an NVSS counterpart, significantly larger than for both the SC and FBC catalogs. This suggests that the SPAM catalog is much less contaminated by false detections than the SC and FBC catalogs, resulting in a larger absolute number of true detections.

This is further strenghtened by the results from the real J1300-208 data set. For this test case, the SPAM image has the largest number of source detections. Again, the SPAM catalog has the largest fraction of associations with the NVSS catalog, the same fraction as with the J0900+398 data set. In contrast, the fraction of NVSS counterparts for SC and VLSS have both gone down. This is best explained by an increase in (non-Gaussian) sidelobe noise in the image background due to calibration errors, which corresponds with the observed increase in $\sigma$.

\subsubsection{Source Peak Fluxes}
\label{sec:src_photo}

\begin{figure*}
\begin{center}
\resizebox{0.8\hsize}{!}{
\includegraphics[angle=0]{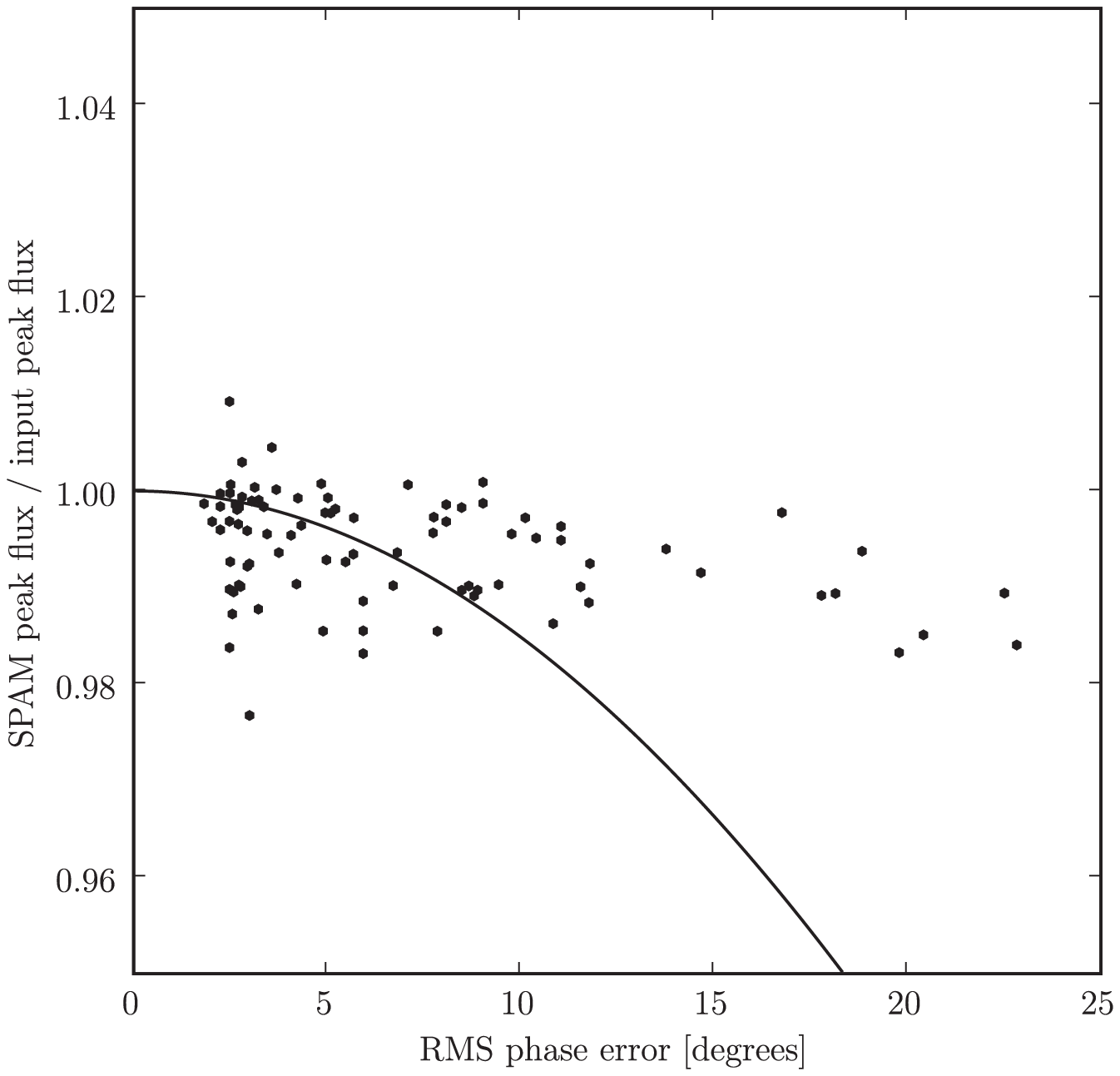}
\includegraphics[angle=0]{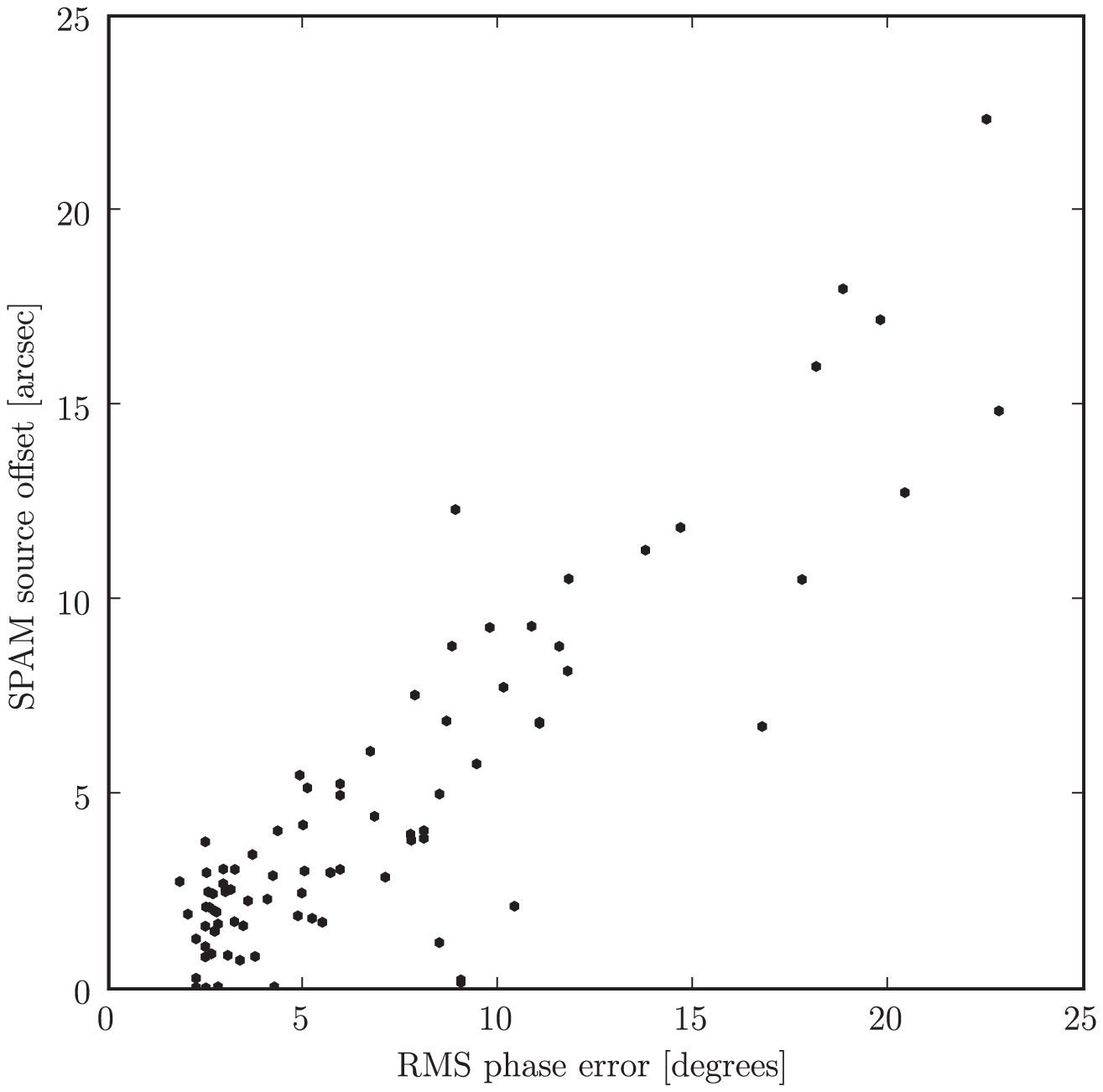}}
\caption{Peak flux ratios in the simulated J0900+398 field: \textit{Left}: Peak flux ratios of the 91~extracted sources from the SPAM image as compared to the input model sources, plotted as a function of the residual RMS phase error after SPAM calibration. Overplotted is the theoretical Strehl ratio (solid line) as given in Equation~\ref{eq:strehl_ratio}. For larger RMS phase errors, the measured peak flux ratios do not follow the theoretical strehl ratio curve. This indicates that systematic phase errors dominate the larger RMS phase errors. \textit{Right}: Same peak flux ratios plotted as a function of absolute position offset between extracted sources in the SPAM image and the input model (see Figure~\ref{fig:model_spam_offsets}). The presence of a strong correlation indicates that residual phase gradients dominate the larger RMS phase errors.}
\label{fig:model_phase_errors}
\end{center}
\end{figure*}

The presence of residual phase errors after calibration can cause an unresolved source shape to deviate from a point source shape. The source flux is redistributed over a larger area and the peak flux of the source drops. At 80\arcsec{} resolution, most sources in a VLSS field are unresolved. Therefore, a mean increase of source widths over the point source width is a direct measure of ionospheric conditions. This argument was used in the pre-selection of data sets for our test cases.

For significant source deformations or low SNR sources, determination of the shape of individual sources is subject to large uncertainties (e.g., Condon et al. \cite{bib:condon1997}). Because determination of peak fluxes is much more robust, we use these for a relative comparison of calibration accuracy. Starting with the original catalogs as produced by BDSM, we associate sources between the undisturbed, FBC, SC and SPAM catalogs that lie within 80\arcsec{} of the same NVSS source and has a peak flux larger than $5\,\sigma$ in at least one of the two catalogs.

For the simulated J0900+398 data set, the true peak fluxes of all 91 sources are known. A comparison between peak fluxes from the undisturbed image and the input catalog identifies a small ($< 1$ percent) CLEAN bias of 3.6~\mjybeam (e.g., Condon et al. \cite{bib:condon1994}, \cite{bib:condon1998}; Becker, White \& Helfand \cite{bib:beckerwhitehelfand1995}). Ignoring the image noise dependency of CLEAN bias, we applied this small correction to the peak fluxes in the undisturbed, SC and SPAM source catalogs before proceeding. Figure~\ref{fig:model_spam_sc_peak_peak} shows a comparison of the measured-to-input peak flux ratios for sources in the SC and SPAM images. The mean peak flux ratio for both images is approximately equal and just slightly smaller than one. The larger scatter in the SC peak fluxes is consistent with a higher $\sigma$. Using Equation~\ref{eq:strehl_ratio}, the random part of the mean RMS phase error for both SC and SPAM is estimated at 5--6~degrees. This value is comparable to the observed RMS phase error over large parts of the SPAM image (Section~\ref{sec:phase}).

To study the nature of residual RMS phase errors after application of SPAM, we plot the RMS phase errors at the source positions from Figure~\ref{fig:model_sources_errors} against SPAM-to-input peak flux ratios (Figure~\ref{fig:model_phase_errors}). For Gaussian random phase errors, the peak flux ratio is expected to decrease with increased RMS phase error as described in Equation~\ref{eq:strehl_ratio}. However, the discrepancy between the data points and Equation~\ref{eq:strehl_ratio} indicates that for larger RMS values the phase errors are predominantly systematic rather than random.

\begin{figure*}
\begin{center}
\resizebox{\hsize}{!}{
\includegraphics[width=0.1\hsize,angle=0]{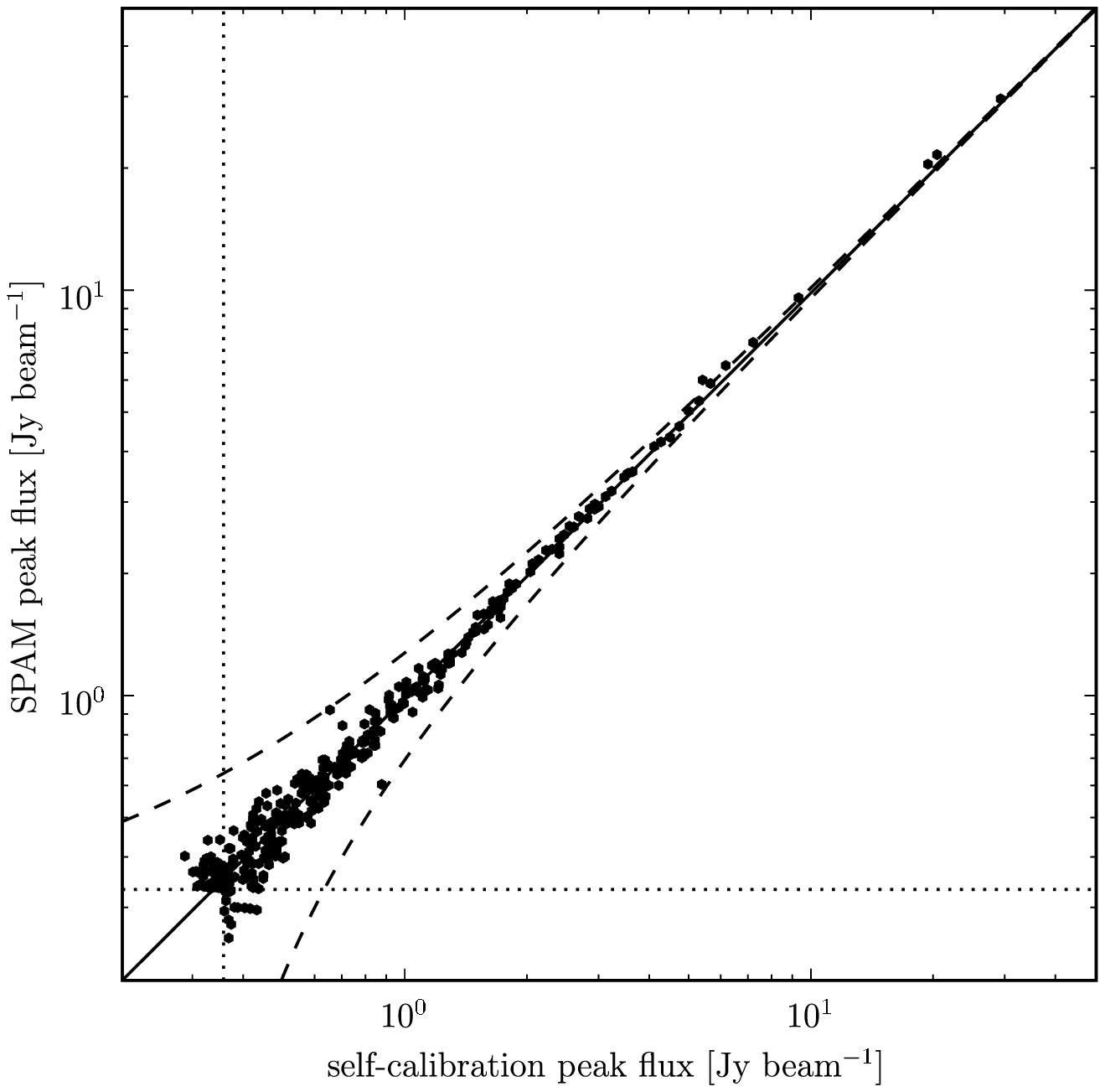}
\includegraphics[width=0.1\hsize,angle=0]{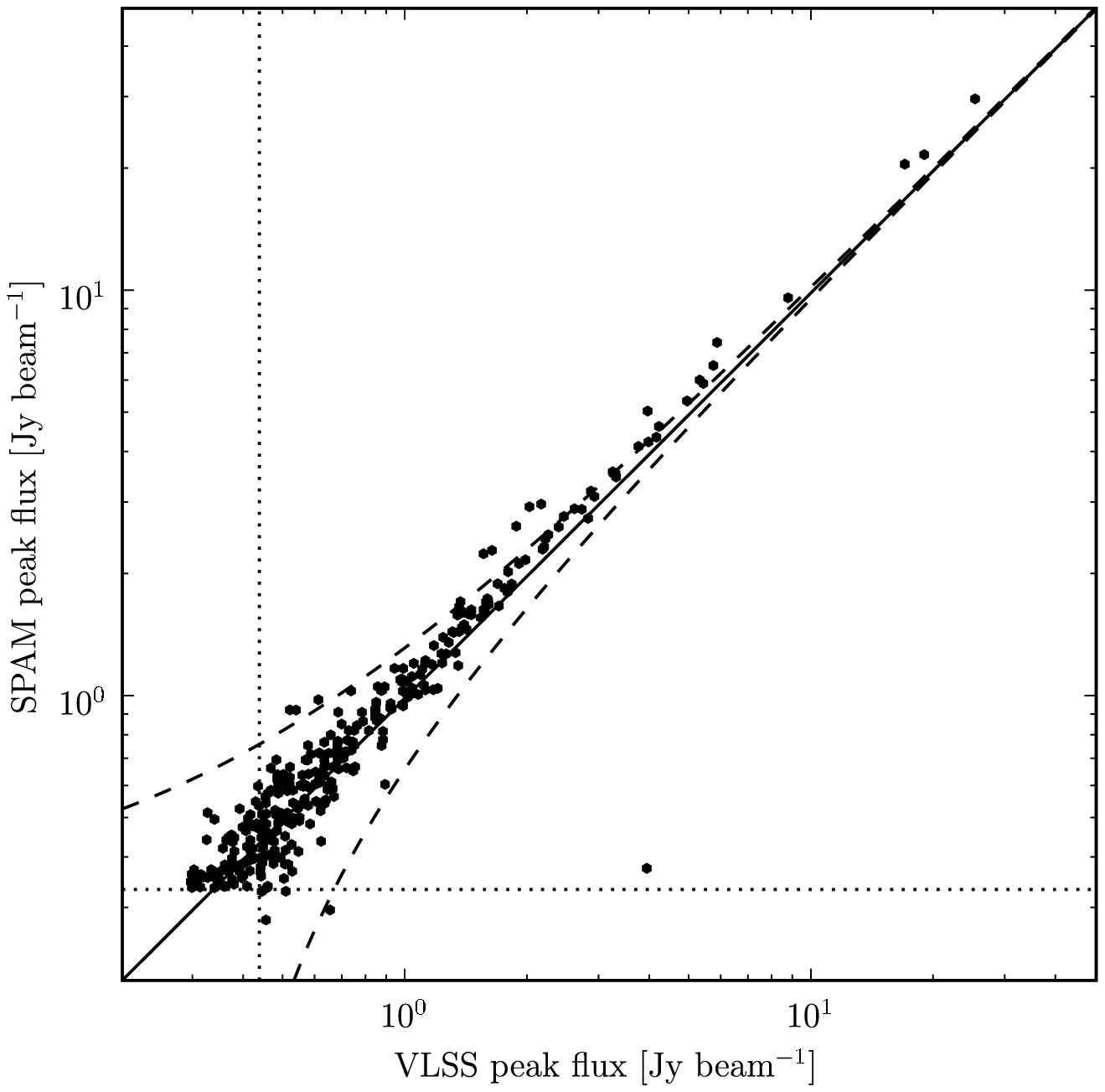}
\includegraphics[width=0.1\hsize,angle=0]{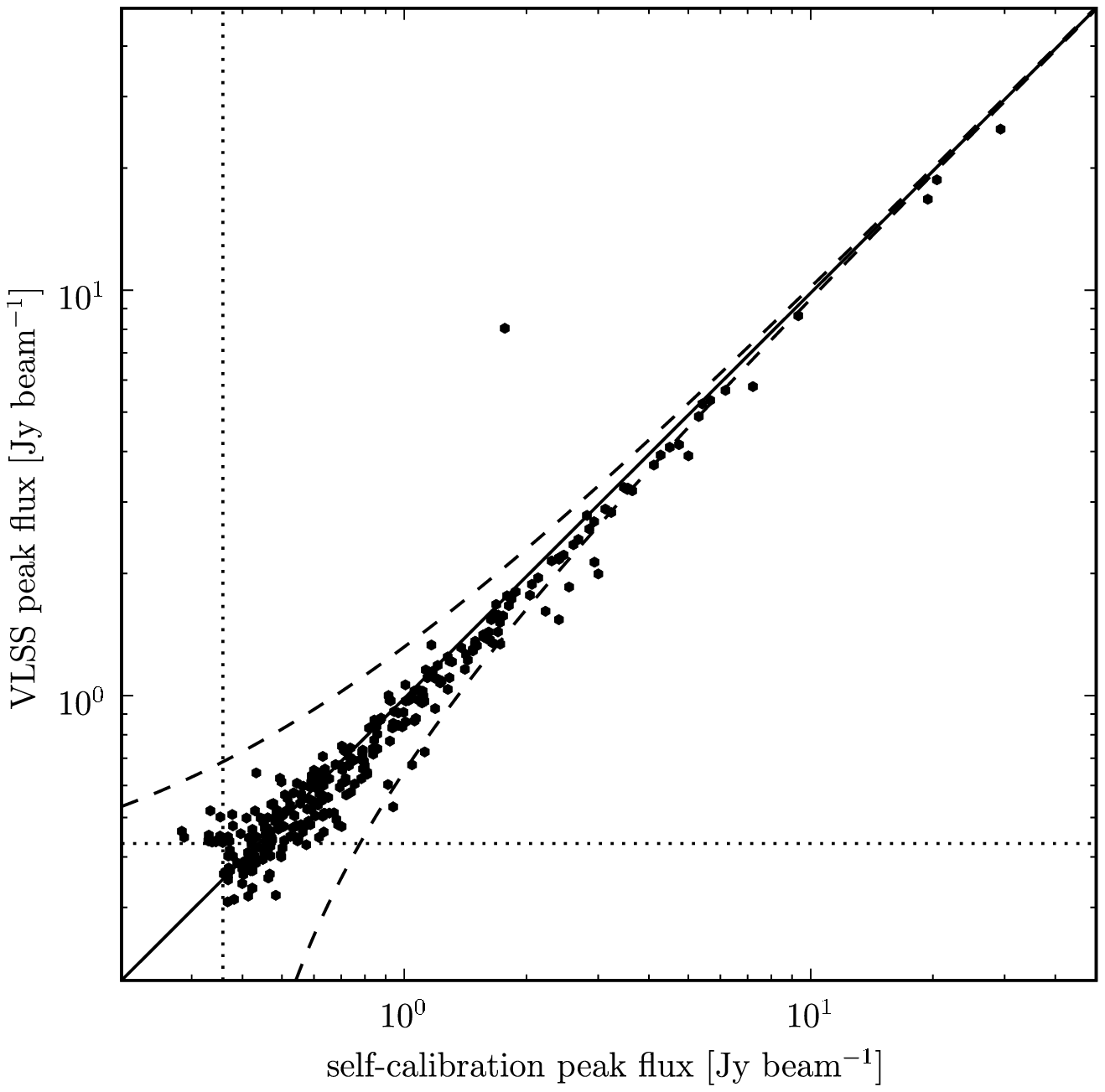}}
\caption{Peak fluxes in the real J0900+398 field \textit{Left}: Peak flux comparison for 367~sources detected in both the self-calibration and SPAM images. The straight diagonal line represents equality, the dashed lines represent $3\,\sigma^{}_\mathrm{C}$ deviations (where $\sigma^{}_\mathrm{C}$ is the combined noise level from both images), and the dotted lines indicate the $5\,\sigma$ detection limit. For bright sources (peak fluxes $\gtrsim 1$~\jybeam), the average peak flux ratio is 1.00. \textit{Middle}: Same for 329 sources in the field-based calibration (VLSS) and SPAM images. The average bright peak flux ratio of SPAM over field-based calibration is 1.10. \textit{Right}: Same for 313 sources in the self-calibration and field-based calibration (VLSS) images. The average bright peak flux ratio of self-calibration over field-based calibration is 1.10. In all plots, the image noise causes a larger scatter in the peak flux determinations of faint sources ($\lesssim 1$~\jybeam) and consequently, a selection bias towards positively enhanced peak fluxes that increases with image noise.}
\label{fig:0900_peak_peak}
\end{center}
\end{figure*}

\begin{figure*}
\begin{center}
\resizebox{\hsize}{!}{
\includegraphics[width=0.1\hsize,angle=0]{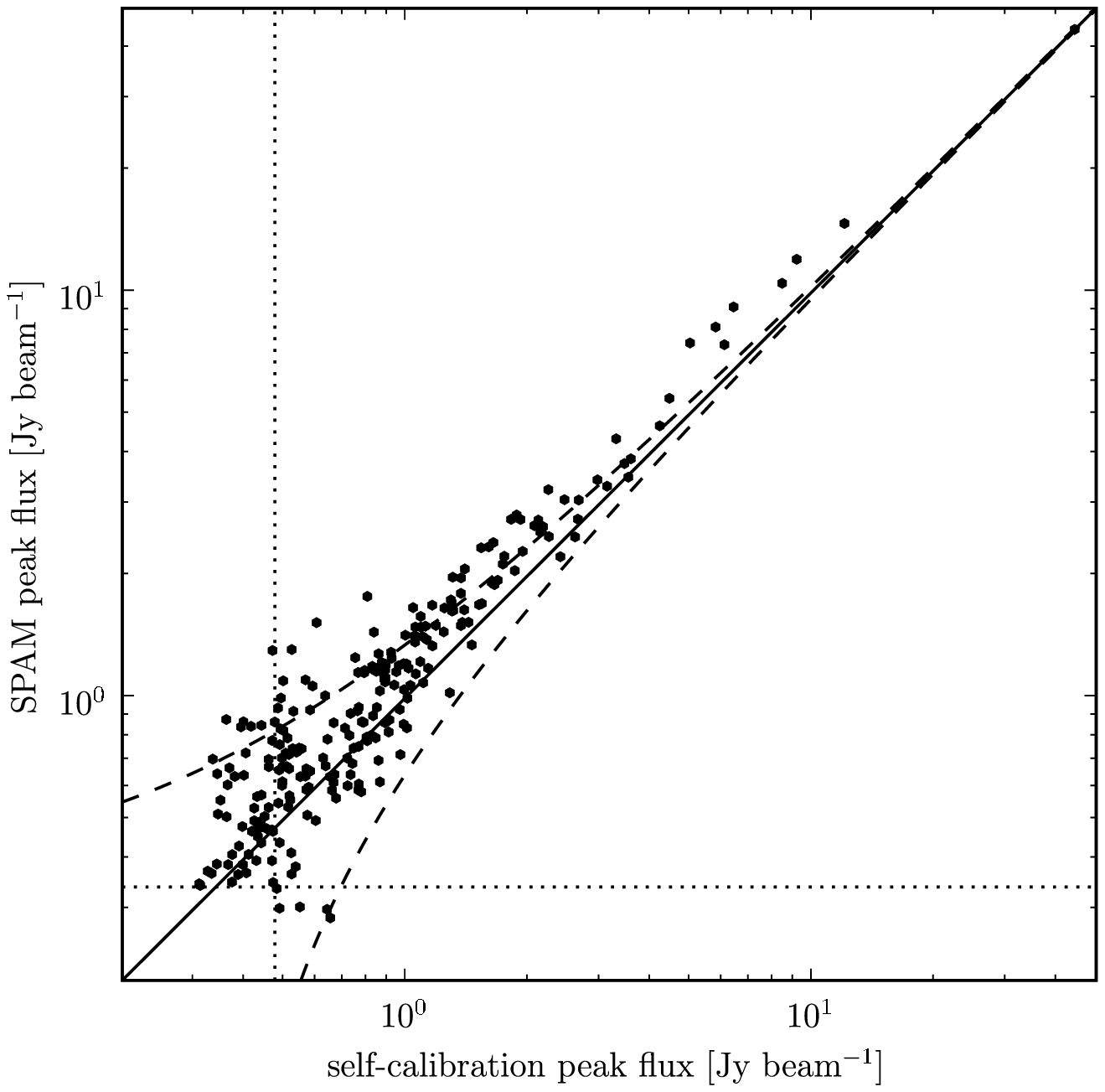}
\includegraphics[width=0.1\hsize,angle=0]{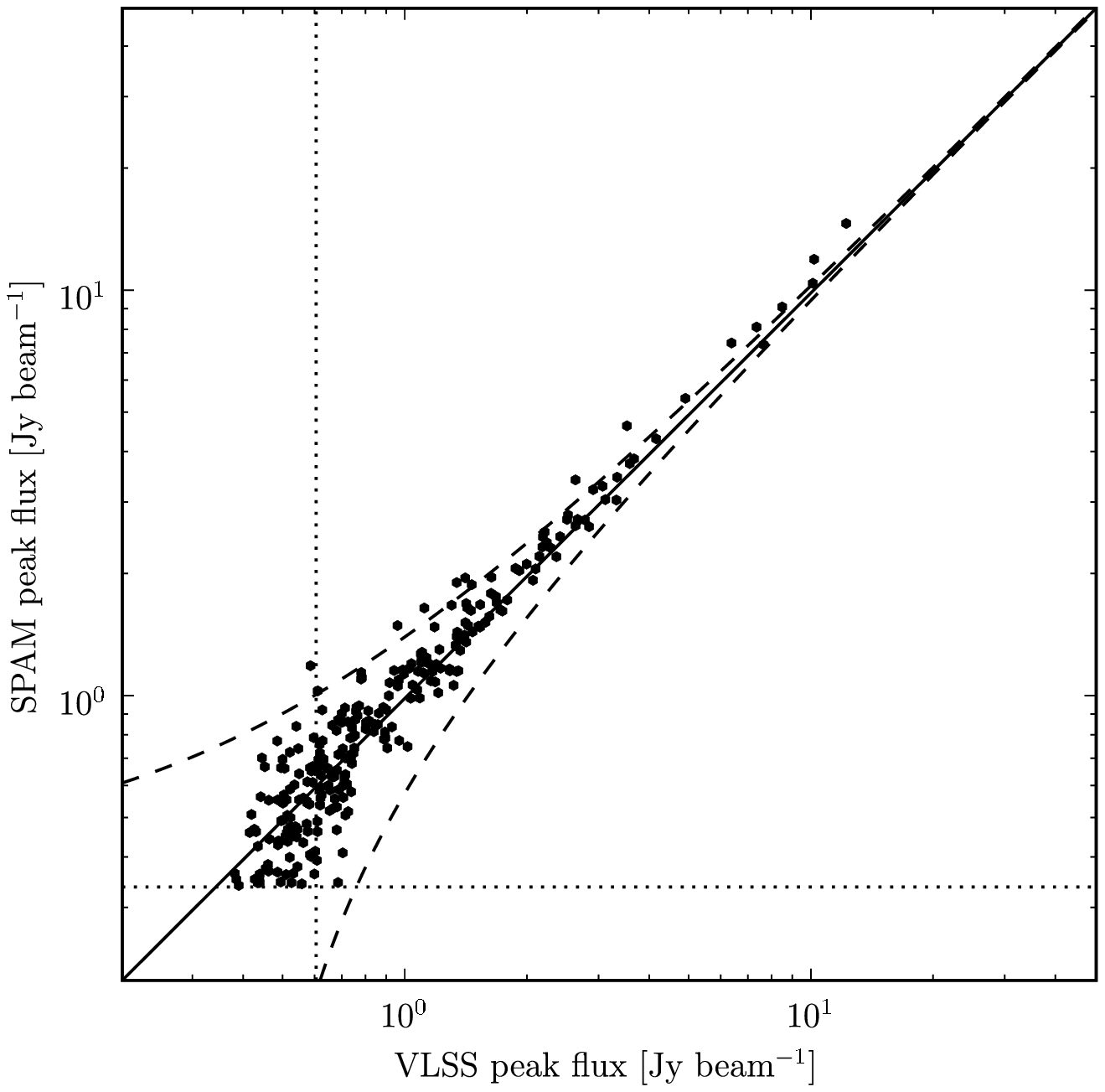}
\includegraphics[width=0.1\hsize,angle=0]{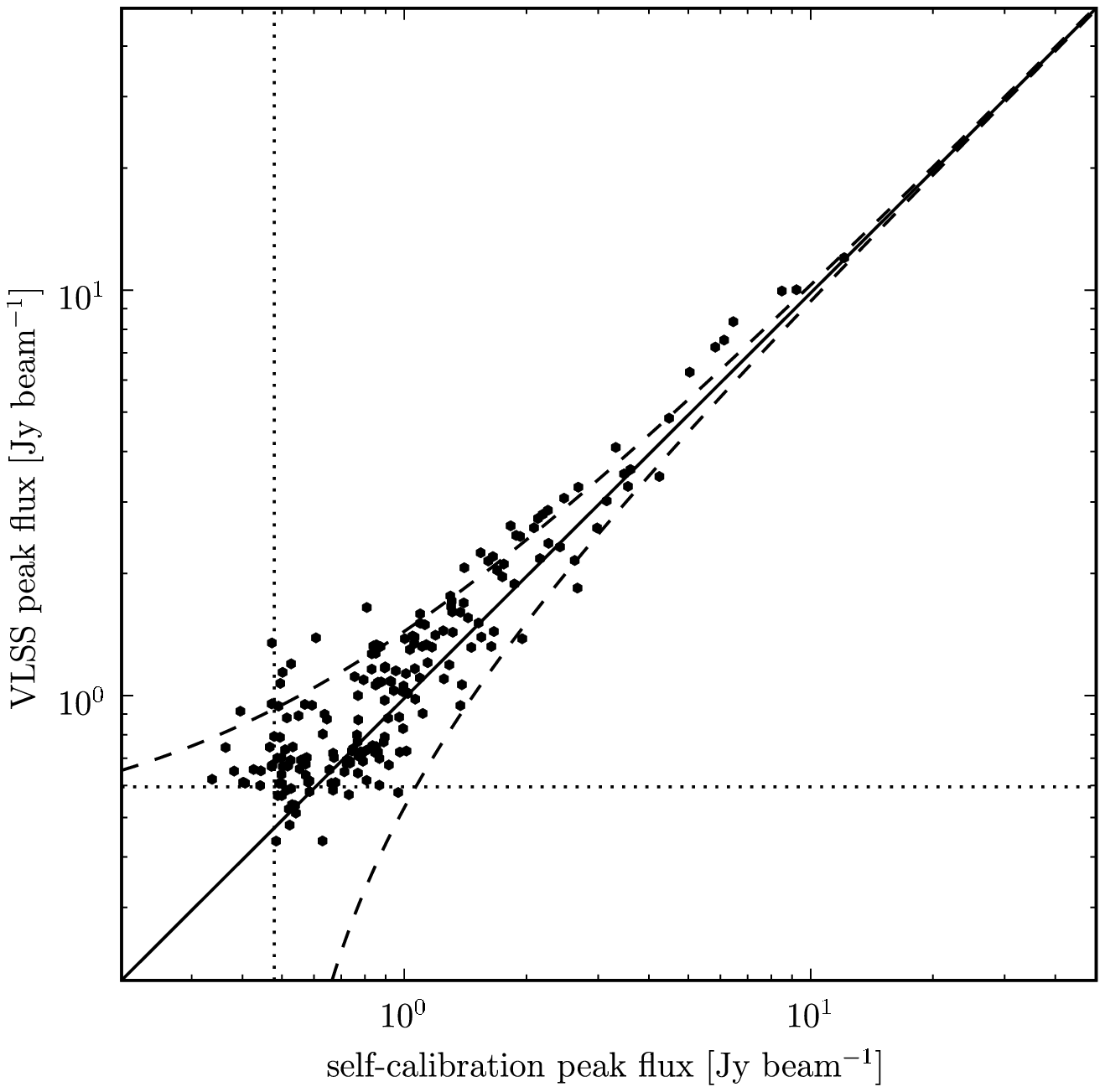}}
\caption{Peak fluxes in the (real) J1300-208 field: \textit{Left}: Peak flux comparison for 247 sources detected in both the self-calibration and SPAM images. For bright sources (peak fluxes $\gtrsim 1$~\jybeam), the average peak flux ratio of SPAM over SC is 1.24. \textit{Middle:} Same for 278 sources in the field-based calibration (VLSS) and SPAM images. The average bright peak flux ratio of SPAM over field-based calibration is 1.07. \textit{Right:} Same for 202 sources in the self-calibration and field-based calibration (VLSS) images. The average bright peak flux ratio of field-based calibration over self-calibration is 1.15.}
\label{fig:1300_peak_peak}
\end{center}
\end{figure*}

For the real J0900+398 data set, Figure~\ref{fig:0900_peak_peak} shows a comparison of peak fluxes for associated sources in the SC, FBC and SPAM catalogs. There is a good match between peak fluxes measured in the SC and SPAM catalogs. For high SNR sources with a peak flux above 1~Jy, the SPAM peak fluxes match on average within 1~percent with the SC peak fluxes. Similarly, SC and SPAM peak fluxes are on average 10~percent higher than FBC peak fluxes. The systematic increase of peak fluxes for SC and SPAM as compared to FBC for many more than the calibrator sources denotes a more accurate calibration over large parts of the FoV. Towards the low flux end, source detections are slightly biased towards the image with the highest noise level, which is the FBC image.

Figure~\ref{fig:1300_peak_peak} shows the same comparison of peak fluxes for the real J1300-208 data set. For high SNR sources with a peak flux above 1~Jy, the SC peak fluxes are by far the smallest, while FBC and SPAM peak fluxes are on average higher by 15 and 24~percent, respectively. The relative loss of peak flux in the SC image is a clear indication of the break-down of the assumption of isoplanaticity across the FoV. Under the conditions that clearly need direction-dependent corrections, the SPAM peak fluxes are on average 7~percent higher than the FBC peak fluxes.

\subsubsection{Astrometry}
\label{sec:src_astro}

\begin{figure*}
\begin{center}
\resizebox{0.8\hsize}{!}{
\includegraphics[width=0.1\hsize,angle=0]{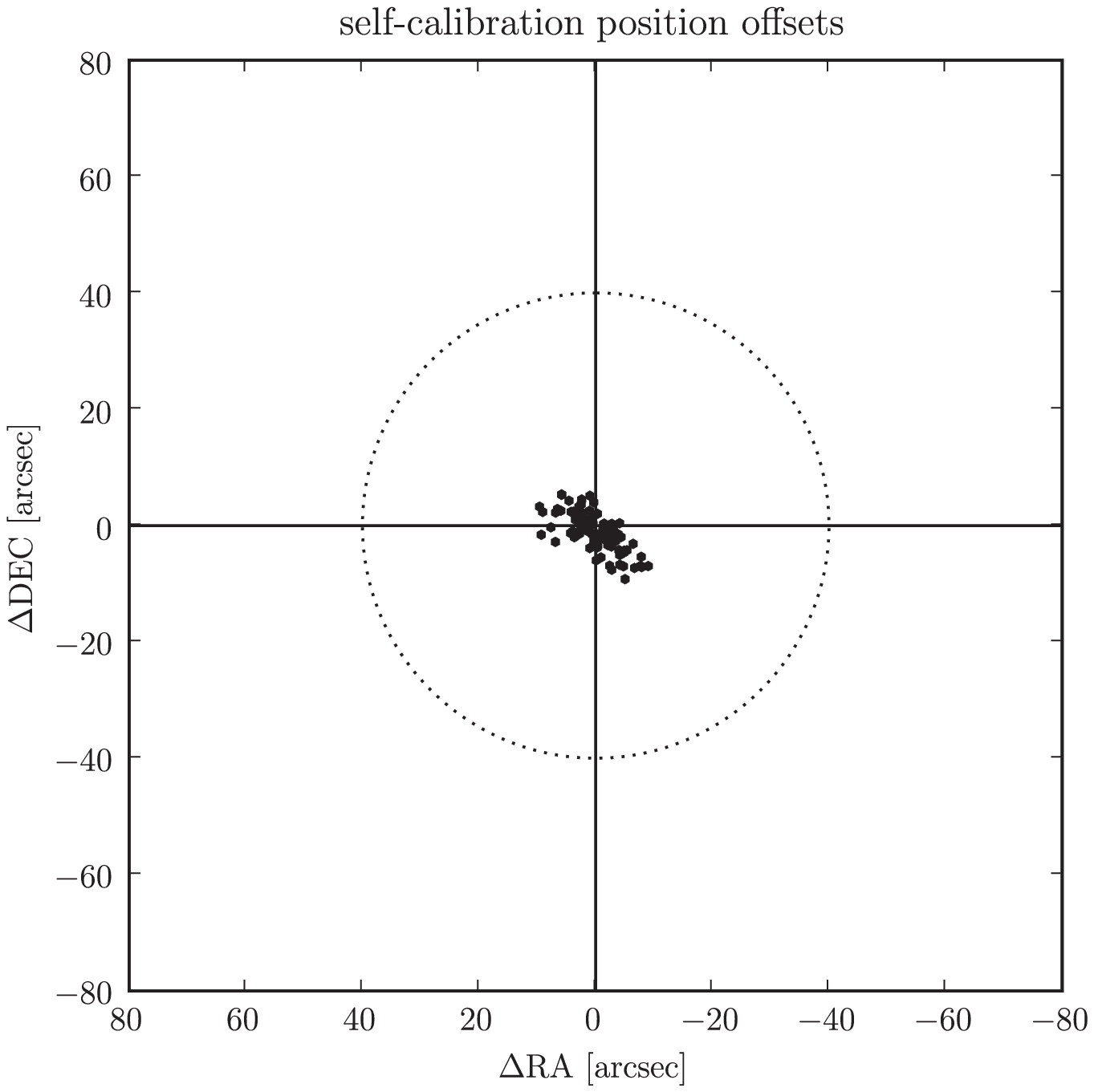}
\includegraphics[width=0.1\hsize,angle=0]{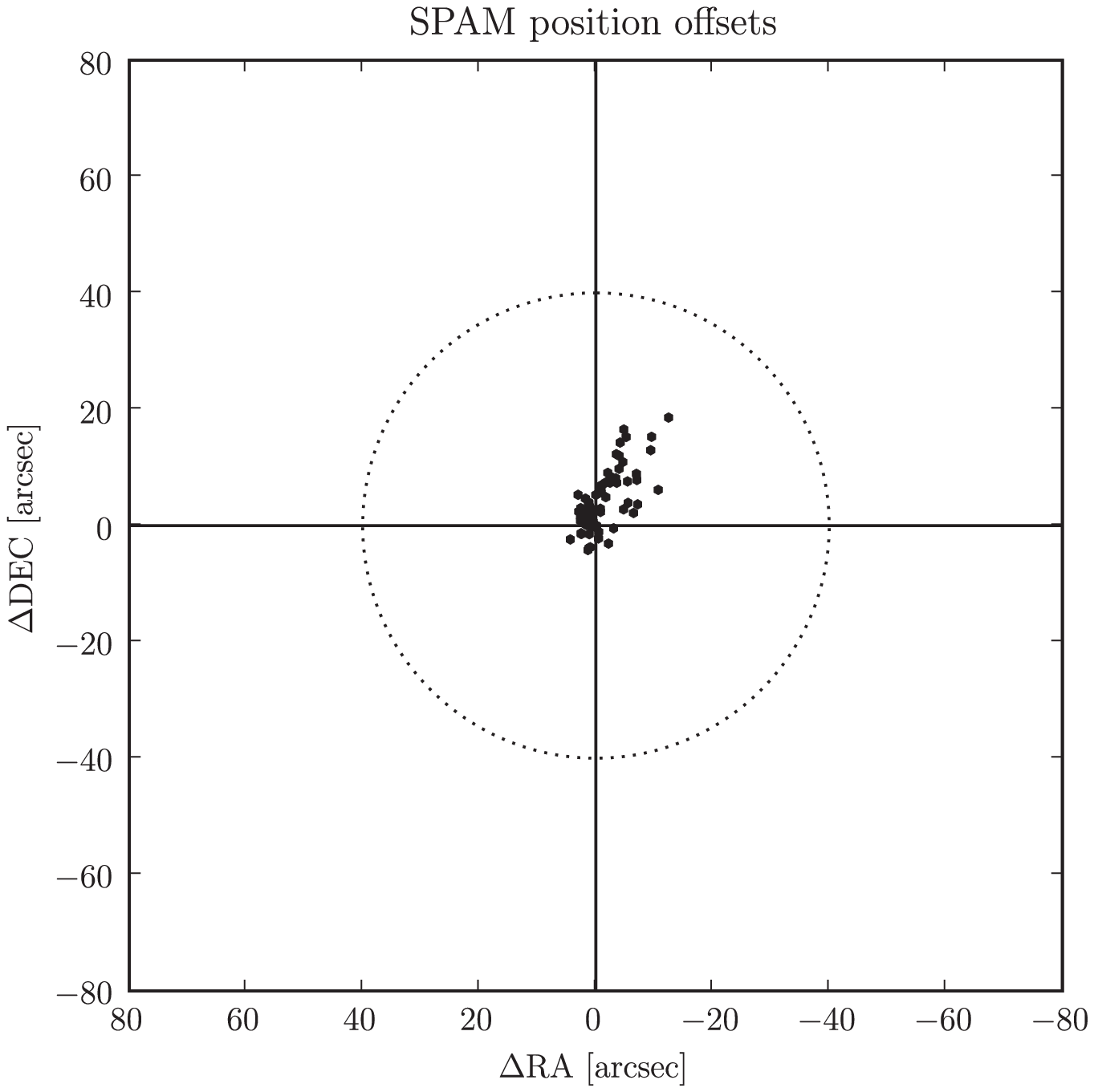}}
\caption{Position offsets in the simulated J0900+398 field: \textit{Left}: Offsets between the measured source positions in the self-calibration image as compared to the input model. \textit{Right}: Same for the SPAM image. In both cases, the distribution around the origin is non-Gaussian. For the SPAM image, the tail of points extending roughly northwards indicates the presence of persistent phase gradients in local parts of the SPAM image. All source position offsets fall well within the size of the 80\arcsec{} restoring beam (dotted line).}
\label{fig:model_spam_offsets}
\end{center}
\end{figure*}

When the time-average of residual phase errors towards a source contains a non-zero spatial gradient, the source will appear to have shifted its position in the final image (see Section~\ref{sec:img_effects}). This gradient may indicate a limitation of the calibration model to reproduce the ionospheric phase corruptions (e.g., in the absence of nearby calibrators), but may also be introduced by the peeling process. The latter occurs when a peeling source is re-centered to the wrong catalog position (see Section~\ref{sec:peeling}). Because such an error propagates into the calibration model, many sources in the vicinity of the peeling source may also suffer from a systematic astrometric error.

For the simulated data set, the peak positions of sources as determined by BDSM were compared against the positions of counterparts in the input model. For the real data sets, we compared against the NVSS catalog instead. When comparing against NVSS positions, apparently large position offsets may occur due to resolution differences and spectral variation across the source. Averaged over a large number of sources, these offsets should have no preferential orientation. In contrast, a residual phase gradient in a certain viewing direction is expected to cause systematic offsets for groups of sources in a certain preferential direction.

For the simulated J0900+398 data set, Figure~\ref{fig:model_spam_offsets} shows that the positions for both SC and SPAM are accurate to within $\sim 10$\arcsec, except for a small tail of $\sim 15$~SPAM sources that have somewhat larger offsets. These sources are all positioned near the edge of the FoV, where the RMS phase error is large (Figure~\ref{fig:model_sources_errors}). Figure~\ref{fig:model_phase_errors} also confirms this by the clear correlation between RMS phase error and absolute position offsets.

\begin{figure*}
\begin{center}
\resizebox{\hsize}{!}{
\includegraphics[width=0.1\hsize,angle=0]{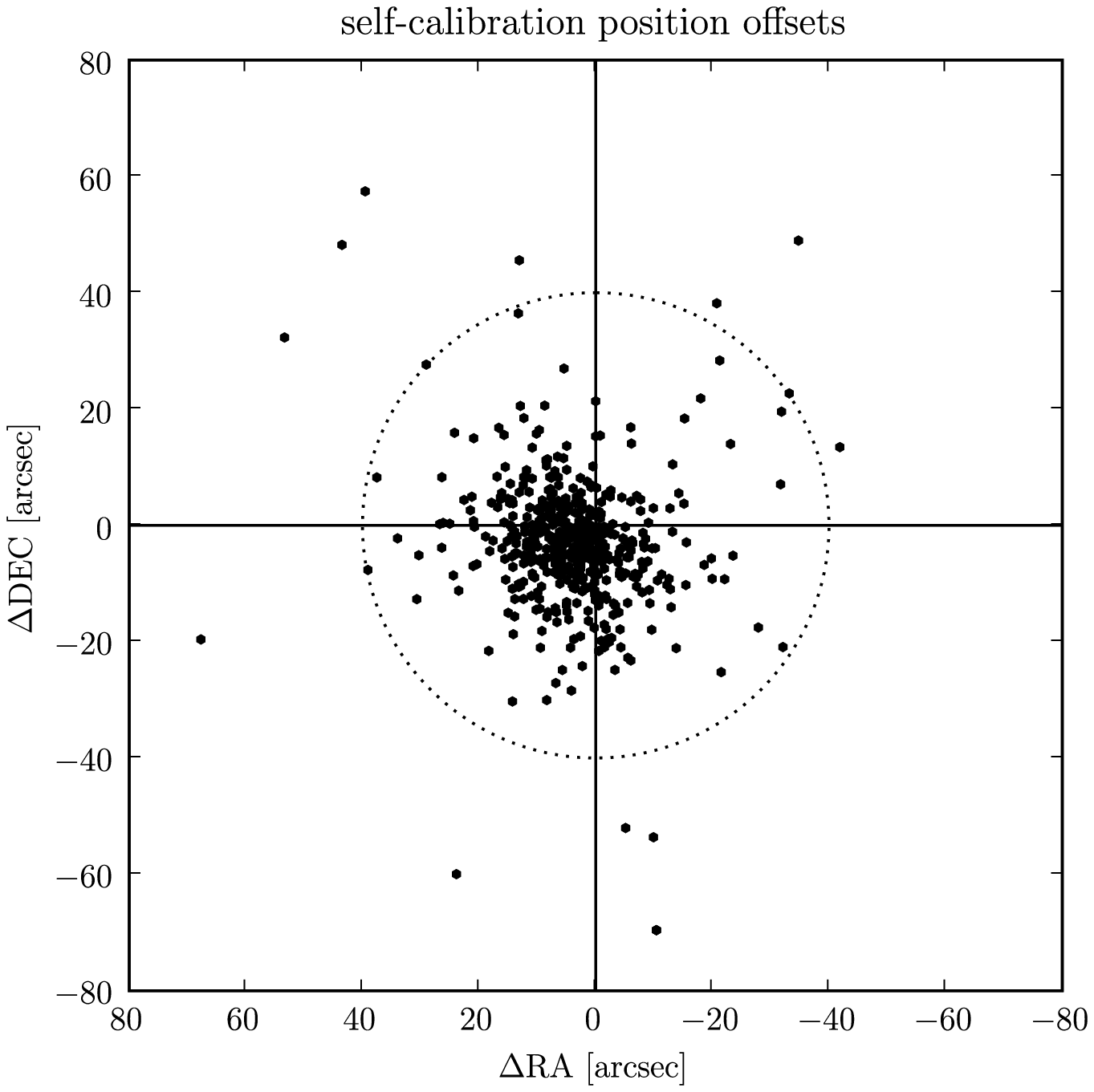}
\includegraphics[width=0.1\hsize,angle=0]{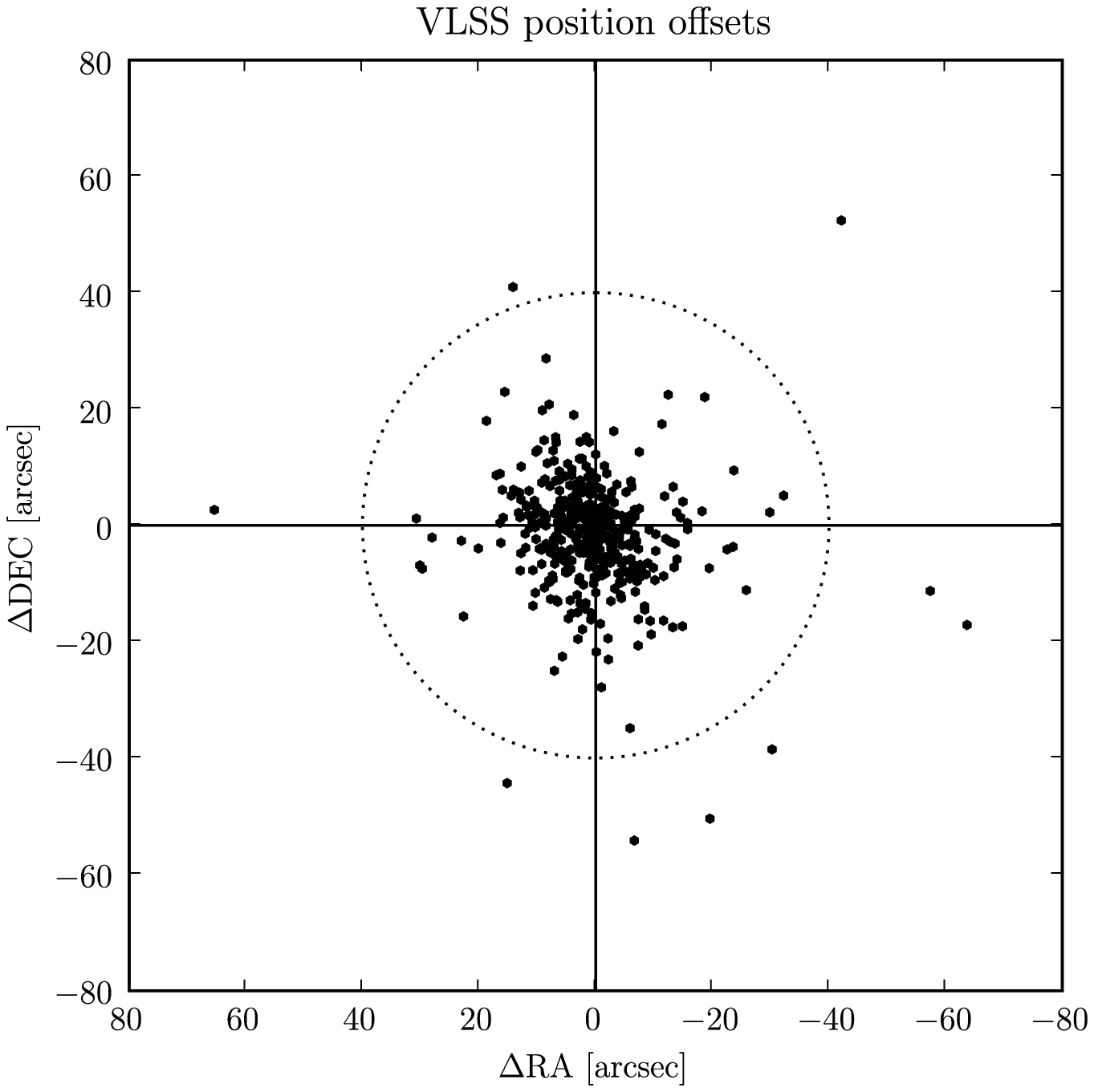}
\includegraphics[width=0.1\hsize,angle=0]{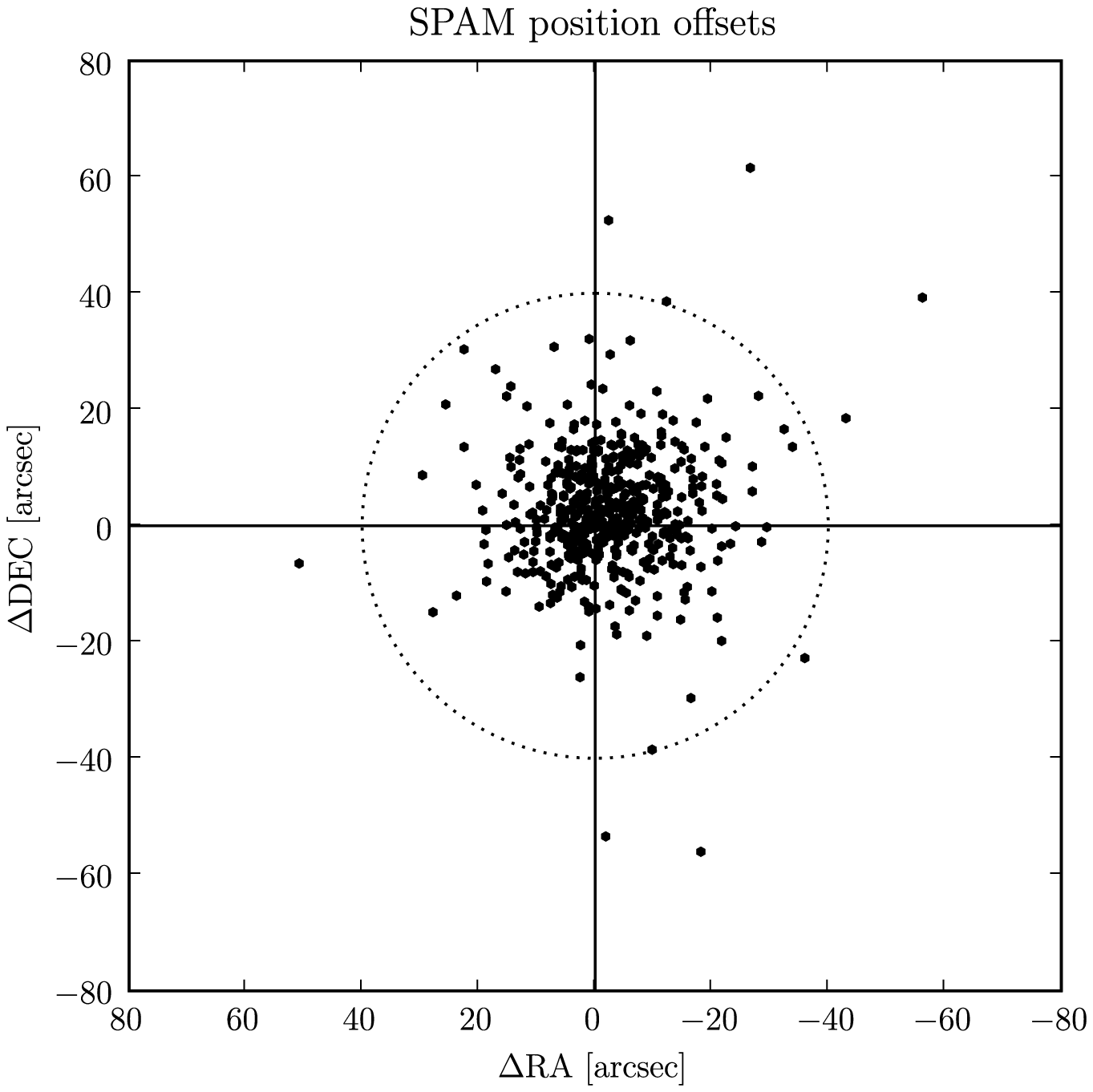}}
\caption{Position offsets in the real J0900+398 field \textit{Left}: Offsets between the measured source positions in the self-calibration image as compared to the NVSS catalog. \textit{Middle}: Same for the field-based calibration (VLSS) image. \textit{Right}: Same for the SPAM image.}
\label{fig:0900_vlss_spam_offsets}
\end{center}
\end{figure*}

For the real J0900+398 data set, the source position offsets for SC, FBC and SPAM relative to NVSS catalog positions are plotted in Figure~\ref{fig:0900_vlss_spam_offsets}. The larger scatter as compared to the simulated J0900+398 data set can be the (combined) result of less accurate position measurements due to higher image noise, resolution and spectral differences between the observations and the NVSS catalog or larger residual RMS phase errors after calibration. The observed scatter for SC is centered around a point that is offset from the origin by $\sim 5$\arcsec, which is either caused by inaccuracies in the initial sky model or during the self-calibration process (Section~\ref{sec:init_sky_calib}). The scatter of both FBC and SPAM offsets is centered close to the origin. The RMS of the scatter around the mean position offset is 10.5\arcsec{} for both FBC and SPAM (despite the apparently larger scatter for SPAM, which is due to a larger number of data points), both smaller than the 11.9\arcsec{} for SC. 

\begin{figure*}
\begin{center}
\resizebox{\hsize}{!}{
\includegraphics[width=0.1\hsize,angle=0]{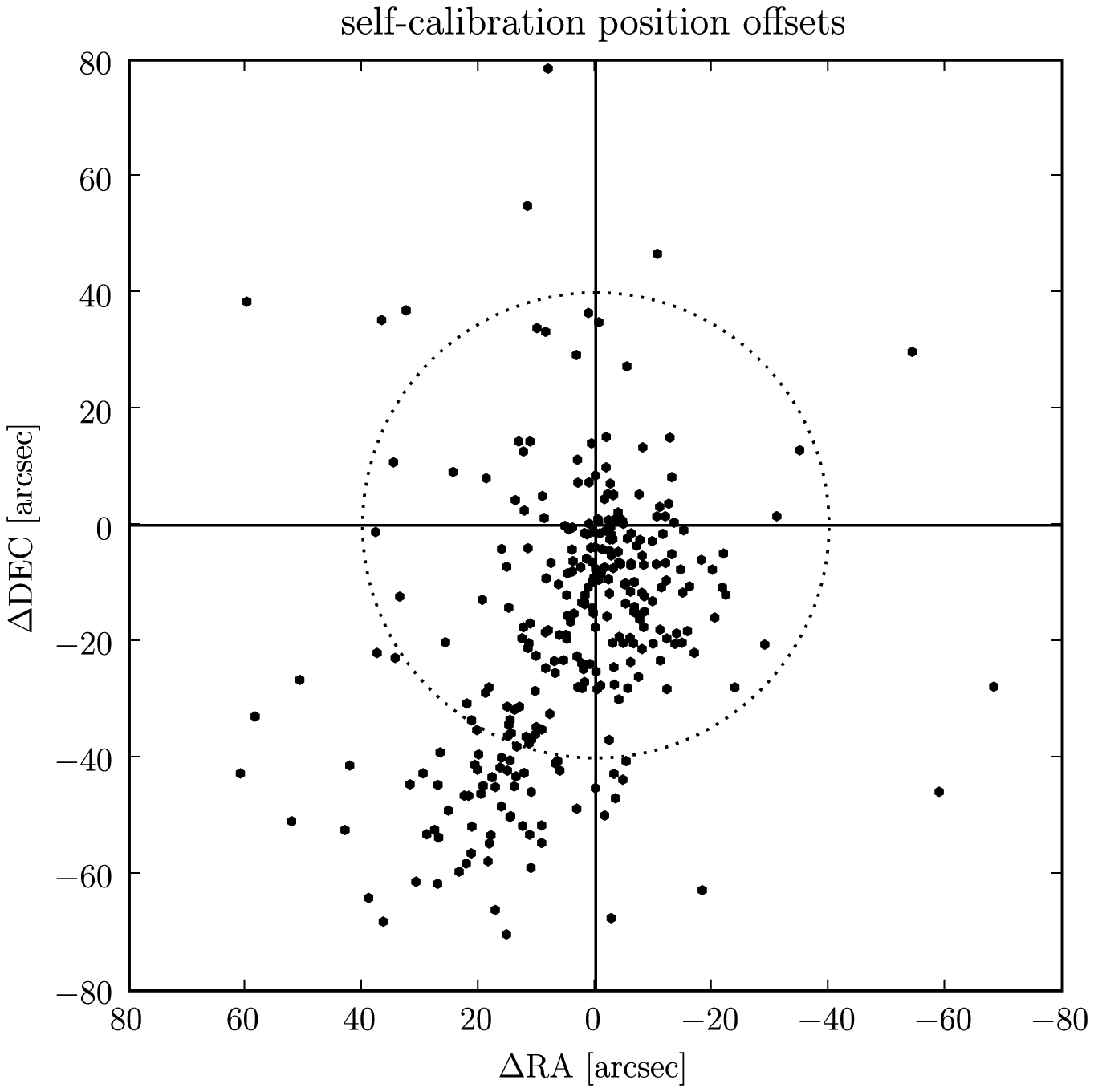}
\includegraphics[width=0.1\hsize,angle=0]{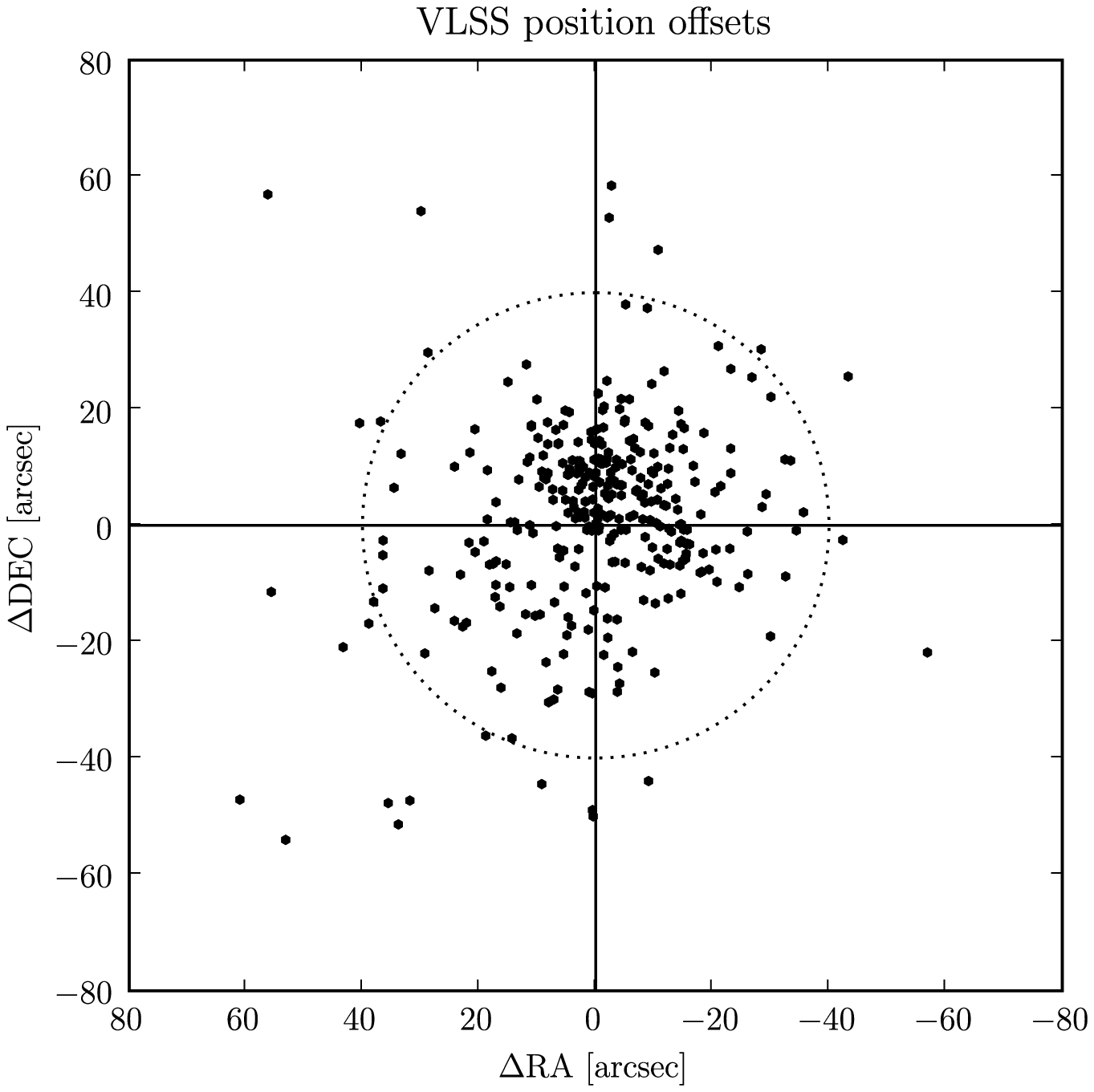}
\includegraphics[width=0.1\hsize,angle=0]{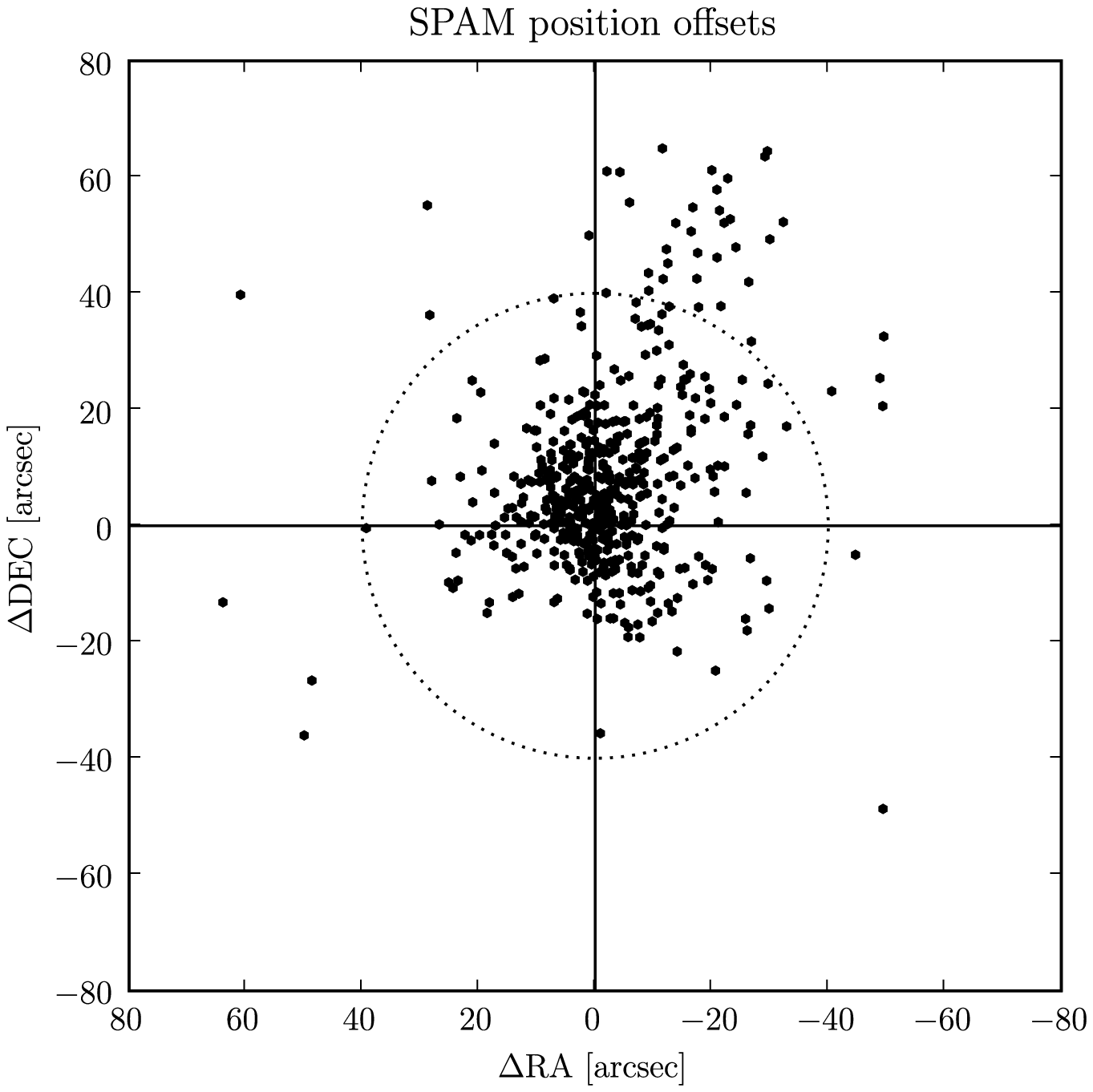}}
\caption{Position offsets in the real J1300-208 field: \textit{Left}: Offsets between the measured source positions in the self-calibration image as compared to the NVSS catalog. \textit{Middle}: Same for the field-based calibration (VLSS) image. \textit{Right}: Same for the SPAM image.}
\label{fig:1300_vlss_spam_offsets}
\end{center}
\end{figure*}

For the real J1300-208 data set, the source position offsets for SC, FBC and SPAM relative to NVSS catalog positions are plotted in Figure~\ref{fig:1300_vlss_spam_offsets}. The position scatter for all three methods is significantly larger than for the real J0900+398 data set, and all suffer from systematic position offsets in varying degrees of severity. The position offsets in the SC image have a seriously distorted distribution, which includes a large tail of points that extends roughly southwards. This indicates the presence of varying systematic source offsets over the whole FoV. The distribution of position offsets in the FBC image is more compact but also asymmetric, and is approximately centered around a point that is $\sim 10$\arcsec{} offset in northward direction from the origin. A large number of the SPAM position offsets are clustered near the origin, similar to the real J0900+398 data set, but there is an additional tail of points that runs roughly northwards. The RMS of the scatter around the mean position offset is 20.7\arcsec, 16.5\arcsec{} and 14.8\arcsec{} for SC, FBC and SPAM, respectively, which confirms the apparently strongest clustering of points in the SPAM position offset plot.

Systematic position offsets in the images can be reduced by distortion and regridding of the images. To this purpose, Cohen et al. (\cite{bib:cohen2007}) fit a fourth order Zernike polynomial to the (time constant) position offsets of typically more than 100~sources in the FBC images of the VLSS. They estimate that, after correction, the final residual position error in the full VLSS catalog due to the ionosphere is $\lesssim 3$\arcsec{} in both RA and DEC. 

\section{Discussion and Conclusions}
\label{sec:concl}

The SPAM method for ionospheric calibration has been succesfully tested on one simulated and two carefully selected visibility data sets of 74~MHz observations with the VLA (taken from the VLSS; Cohen et al.~\cite{bib:cohen2007}). From the results of these test cases, we draw the following conclusions: \\
(i) A proof-of-concept is given for several different techniques that were incorporated in SPAM calibration. The peeling technique (Noordam~\cite{bib:noordam2004}) was succesful in providing relative measurements of ionospheric phase errors in the direction of several bright sources in the FoV. The Karhunen-Lo\`eve phase screen (van der Tol \& van der Veen \cite{bib:vdtolvdveen2007}) at fixed height was able to combine these measurements into a consistent model per time stamp. For relatively bad ionospheric conditions, it was demonstrated that the ionospheric calibration cycle (repeated ionospheric calibration and subsequent imaging; Noordam~\cite{bib:noordam2004}) converges within a few iterations to a calibration of similar accuracy as under relatively good ionospheric conditions (for which one iteration was sufficient). \\
(ii) Ionospheric calibration with SPAM is more accurate than the existing self-calibration (e.g., Pearson \& Readhead~\cite{bib:pearsonreadhead1984}) and field-based calibration (Cotton et al.~\cite{bib:cotton2004}) techniques. Even for relatively compact array configuration like VLA-B and BnA, significant improvements in image quality are obtained by allowing for higher-order (i.e., more than a gradient) spatial phase corrections over the array in any viewing direction. In the resulting images, we obtained dynamic range improvements of 5--45\% and 70--80\% under relatively good and bad ionospheric conditions, respectively. \\
(iii) Although the mean astrometric accuracy of source positions in SPAM images is similar to or better than for self-calibration and field-based calibration, systematically larger astrometric errors are present in regions of the output images of all calibration methods. This is caused by a shortage of available calibrators in these regions and positional inaccuracies in the reference source catalog used for calibration.

The 65~\mjybeam noise levels in the SPAM images match the lowest noise levels of the more than 500~images that define the VLSS survey. A potential reduction of the average noise level from 100~\mjybeam to 65~\mjybeam for the full VLSS survey would significantly increase the number of source detections from $\sim$70,000 to about 120,000 (an increase of $\sim 75$\%), but also it would greatly enhance virtually every science goal. For example, using the radio luminosity function for high-luminosity radio galaxies from Jarvis et al. (\cite{bib:jarvis2001}), the estimated number of detectable HzRGs in the VLSS would increase by 65\%, but also the maximum redshift would increase. For a luminous radio galaxy with luminosity of $2 \times 10^{28}$~W\,Hz$^{-1}$\,sr$^{-1}$ at 74~MHz, the redshift limit would rise from $z = 5.7$ to $z = 6.8$. Another example is the detection and study of cluster radio halos.  Using available halo population models (En\ss{}lin \& R\"ottgering \cite{bib:ensslin2002}; Cassano et al. \cite{bib:cassano2006}), the anticipated noise reduction would roughly double the number of detectable halo systems.

For the VLSS, the estimated theoretical thermal noise level of 35~\mjybeam is still a factor of two lower than the average background noise level of $\sim 65$~\mjybeam in the SPAM images. From inspection of the SPAM images we cannot identify an obvious single cause for this. Therefore, similar to Cohen et al.~\cite{bib:cohen2007}, we expect the remaining excess noise to be the combined result of several different causes, including residual ionospheric phase errors after SPAM calibration, but also residual RFI, collective sidelobe noise from many non-deconvolved sources (too faint or outside the FoV) and variable source amplitude errors (e.g., due to pointing errors and non-circular antenna beam patterns; see Bhatnagar et al.~\cite{bib:bhatnagar2008}).

The SPAM test results indicate that the ionospheric calibration accuracy may be further improved. The typical model fit RMS phase error per antenna of $\sim 20-30$~degrees for real data sets is much larger than the $3$~degrees for the noiseless simulated data set. There are several possible sources of error, either in the peeling phase corrections or the ionospheric phase model. Noise in the visibilities (either thermal or non-thermal), contamination from other sources, inaccuracies in the peeling source model and undersampling of the fastest phase fluctuations are factors that degrade the accuracy of peeling. Also, the ionospheric phase screen model may be a poor representation of reality, either because it is incomplete (e.g., absence of vertical structure) or the fixed model parameters are chosen poorly (e.g., screen height, spectral index of phase fluctuations). Several of these issues will be addressed in future work (Section~\ref{sec:future}).

The potential problems with the peeling technique raises the question whether one should use alternative methods. Apart from the precautions described in Section~\ref{sec:peeling}, we have found little means to improve the accuracy of the peeling process for single sources any further. One unexplored option is to peel sources in groups, e.g. identify isoplanatic patches of sky with a large enough total flux from multiple sources. Two possible alternatives approaches to peeling are: (i) simultaneous self-calibration towards multiple sources in the FoV, or (ii) fitting the ionosphere model directly to the visibilities rather than using peeling as an intermediate step. Although these alternative approaches have not been tested by us in practise, we anticipate little improvement over our current accuracy. Van der Tol, Jeffs \& van der Veen (\cite{bib:vdtol2007}) show that, theoretically, iterative peeling converges to the same solution as simultaneous self-calibration. A direct fit of the ionosphere model to the visibilities is, similar to self-calibration, biased towards accurate solving in the direction of the apparently strongest source in the FoV. Although not conclusive for this approach, tests with SPAM show that using even a moderate flux-based weighting into the ionospheric phase model fitting against peeling phase corrections introduces a strong bias towards the brightest source, while calibration accuracy towards other peeled sources degrades severely.

For the existing and future large low-frequency radio interferometer arrays like VLA-A, GMRT, LOFAR, LWA and SKA, the need for a direction-dependent ionospheric calibration method is evident. Based on the results presented in this paper, it is difficult to draw quantitative conclusions on the achievable calibration accuracy for these arrays. If a SPAM-like calibration algorithm is to be used in a very high signal-to-noise observing regime under quiet to moderate ionospheric conditions, it seems likely that residual RMS phase errors in the order of a few degrees could be achieved, comparable to the SPAM results on the simulated VLSS data set.

When relying on the array itself to provide the necessary measurements to constrain ionospheric correction models, ionospheric calibration requires an array layout and sensitivity that allows for sampling the ionsphere over the array at the relevant spatial scales and time resolution. The spatial sampling is determined by the instantaneous pierce point distribution (or more general, the distribution of lines-of-sight through the ionosphere), which depends on the array layout and the detectable calibrator constellation. For future design of low-frequency arrays, it is recommended to optimize the array layout not just for scientific arguments (in general, centrally dense and sparse outside for good UV coverage), but also for ionospheric calibrability (in general, both uniform and randomized).

\section{Future Work}
\label{sec:future}

To test the robustness and limitations of the method, it is necessary to apply SPAM calibration on a wide variety of data sets at different (low) frequencies, obtained with different arrays under different ionospheric conditions. Our highest priority is to test SPAM on observations from the largest existing LF arrays; the VLA in A-configuration and the GMRT. Data for these tests have been obtained and tests are currently in progress. One important possible limitation is the use of a 2-dimensional phase screen to represent the ionosphere. We plan to expand the SPAM model by including multiple screens at different heights and compare the resulting image properties against the current single screen model.

Another limitation of the current implementation is the absence of restrictions on the time behaviour of the model. Antenna-based peeling phases clearly show a coherent temporal behaviour, which is likely to exist for physical reasons. This could be used to reduce the number of required model parameters and suppress the noise propagation from the peeling solutions. We are currently investigating the possibilities of forcing the SPAM model to be continuous in time.

Several of the authors of this article are currently involved in setting up a simulation framework in which one has full control over the sky emission, ionospheric behaviour and array characteristics when generating artificial low frequency observations. Like in the test case on simulated data presented in Section~\ref{sec:apps}, this allows for direct and quantitative comparison between the distorting ionosphere model and the recovered ionospheric phase model by SPAM. We plan to use this setup to further test optimize SPAM calibration for a broad range of ionospheric conditions.

\begin{acknowledgements}

The authors would like to thank Rudolf Le Poole, Reinout van Weeren, Sridharan Rengaswamy, Amitesh Omar, Mamta Pandey, Oleksandr Usov, James Anderson, Ger de Bruyn, Jan Noordam, Maaijke Mevius, Dharam Vir Lal, C.H.~Ishwara-Chandra, A.~Pramesh Rao, Jim Condon and Juan Uson for useful discussions. Special thanks to Mark Kettenis for fast bug fixing and implementation of new functionality in ParselTongue, and Niruj Ramanujam Mohan for similar performance on BDSM. The authors also thank the anonymous referee for useful comments and suggestions. HTI acknowledges a grant from the Netherlands Research School for Astronomy (NOVA). SvdT acknowledges NWO-STW grant number DTC.5893. This publication made use of data from the Very Large Array, operated by the National Radio astronomy Observatory. The National Radio Astronomy Observatory is a facility of the National Science Foundation operated under cooperative agreement by Associated Universities, Inc.
\end{acknowledgements}

\appendix
\section{Derivation and interpolation of the KL base vectors}
\label{app:kl}

This Section contains an outline of the derivation and interpolation of the Karhunen-Lo\`eve (KL) base vectors that are used to describe the ionospheric phase screen in SPAM. The KL technique is adopted from the work by van der Tol \& van der Veen (\cite{bib:vdtolvdveen2007}). For a given stochastic model of spatial electron density fluctuations in the ionosphere, the differential phase rotation on rays of passing radio waves can be described by a (zero mean) isotropic phase screen $\phi(\vec{p})$ with a given spatial covariance function 
\begin{equation}
C^{}_{\phi \phi}(r) = \langle \phi(\vec{p}) \phi(\vec{p}+\vec{r}) \rangle,
\label{eq:cov_r}
\end{equation}
where $\langle\dots\rangle$ denotes the expected value and $r = |\vec{r}|$ is the length of vector $\vec{r}$. In Kolmogorov turbulence theory, a phase structure function is defined as (Equation~\ref{eq:str_func_body})
\begin{equation}
D^{}_{\phi \phi}(\vec{r}) = \langle [ \phi(\vec{p}) - \phi(\vec{p}+\vec{r}) ]^2 \rangle.
\label{eq:struc}
\end{equation}
The structure function and the covariance function are related through
\begin{eqnarray}
D^{}_{\phi \phi}(r) & = & 2 \left[ C^{}_{\phi \phi}(0) - C^{}_{\phi \phi}(r) \right] \quad \Leftrightarrow \nonumber \\
C^{}_{\phi \phi}(r) & = & C^{}_{\phi \phi}(0) - \frac{1}{2} D^{}_{\phi \phi}(r),
\label{eq:struc_cov}
\end{eqnarray}
where $C^{}_{\phi \phi}(0) \equiv \sigma^{2}_{\phi}$ is the phase variance. The phase structure function of a plane wave that passed through a turbulent layer is found to behave as a power-law over a large range of spatial scales (see Section~\ref{sec:ionosphere})
\begin{equation}
D^{}_{\phi \phi}(r) = ( r / r^{}_0 )^{\gamma}_{},
\label{eq:phase_structure_function}
\end{equation}
where $r^{}_0$ is a measure of the scale size of phase fluctuations and $\gamma$ is the power-law slope.

Because the domain of $\vec{p}$ is a limited set of $P$ discrete ionspheric pierce points ${\vec{p}}$, we switch to matrix notation. The elements of the ($P \times P$) phase covariance matrix are given by
\begin{equation}
\mathbf{C}^{}_{\phi \phi}[i,j] = C^{}_{\phi \phi}( r^{}_{ij} ),
\label{eq:cov_ij}
\end{equation}
with $\vec{p}^{}_i, \vec{p}^{}_j \in \{\vec{p}\}$ and $r^{}_{ij} = | \vec{p}^{}_i - \vec{p}^{}_j |$. The symmetric covariance matrix can be decomposed into
\begin{equation}
\mathbf{C}^{}_{\phi \phi} = \mathbf{U} \mathbf{\Lambda} \mathbf{U}^\mathrm{T}_{},
\label{eq:cov_decomp}
\end{equation}
where the columns of ($P \times P$) matrix $\mathbf{U}$ contain the orthonormal eigenvectors of $\mathbf{C}^{}_{\phi \phi}$, $\mathbf{U}^\mathrm{T}_{}$ is the transpose of $\mathbf{U}$ and the ($P \times P$) diagonal matrix $\mathbf{\Lambda}$ contains the eigenvalues. The eigenvectors in the columns of $\mathbf{U}$ are a suitable set of base vectors to describe the phase screen at the pierce points $\{\vec{p}\}$. The eigenvalues on the diagonal of $\mathbf{\Lambda}$ are a measure of the variance of the eigenvector coefficients. Reducing the fitting order is an arbitrary, but necessary step. An optimal subset of eigenvectors is determined by selecting only those with the largest eigenvalues, as the coefficients with the highest variances are the most significant in the modeling problem.

When the fitting order is reduced from $P$ to $\tilde{P} < P$, we are left with a subset of eigenvectors in the columns of ($P \times \tilde{P}$) matrix $\mathbf{\tilde{U}}$, and eigenvalues on the diagonal of ($\tilde{P} \times \tilde{P}$) matrix $\mathbf{\tilde{\Lambda}}$, for which
\begin{equation}
\mathbf{C}^{}_{\phi \phi} \approx \mathbf{\tilde{U}} \mathbf{\tilde{\Lambda}} \mathbf{\tilde{U}}^\mathrm{T}_{}.
\label{eq:cov_approx}
\end{equation}
The phase screen at the pierce point locations can be approximated by
\begin{equation}
\mathbf{\Phi} \approx \mathbf{\tilde{U}} \mathbf{q},
\label{eq:phi_approx}
\end{equation}
where we have denoted the phases and eigenvector coefficients in matrix notation as $\mathbf{\Phi}$ ($P \times 1$) and $\mathbf{q}$ ($\tilde{P} \times 1$), respectively. $\mathbf{q}$ is unknown and needs to be solved for by use of a non-linear least squares method using Equation~\ref{eq:chi_squared}. When $\mathbf{q}$ is determined, the phase screen can be evaluated at arbitrary pierce point locations $\{\hat{\vec{p}}\}$ through Kriging interpolation (Matheron~\cite{bib:matheron1973}):
\begin{equation}
\hat{\mathbf{\Phi}} = \mathbf{C}^{}_{\hat{\phi} \phi} {\mathbf{C}^{-1}_{\phi \phi}} \mathbf{\tilde{U}} \mathbf{q}
\label{eq:kriging}
\end{equation}
where $\mathbf{C}^{}_{\hat{\phi} \phi}$ is the ($\hat{P} \times P$) covariance matrix between $\{\hat{\vec{p}}\}$ and $\{\vec{p}\}$. Inversion of the full $\mathbf{C}^{}_{\phi \phi}$ can be approximated using
\begin{equation}
{\mathbf{C}^{-1}_{\phi \phi}} \approx \mathbf{\tilde{U}} \mathbf{\tilde{\Lambda}}^{-1} \mathbf{\tilde{U}}^\mathrm{T}_{}.
\label{eq:icov_approx}
\end{equation}

The elements of the covariance matrix can be calculated using Equations~\ref{eq:struc_cov} and \ref{eq:phase_structure_function}. For our application, the absolute value of $r^{}_0$ is not relevant because we only require the relative eigenvalues for the order reduction. The elements of the ($P \times P$) structure matrix are given by
\begin{equation}
\mathbf{D}^{}_{\phi \phi}[i,j] = D^{}_{\phi \phi}( r^{}_{ij} ).
\label{eq:structure_matrix}
\end{equation}
The relation between the structure matrix and the covariance matrix is given by
\begin{equation}
\mathbf{C}^{}_{\phi \phi} = \sigma^{2}_{\phi} \mathbf{1^{}_{}} \mathbf{1}^\mathrm{T}_{} - \frac{1}{2} \mathbf{D}^{}_{\phi \phi},
\label{eq:structure_covariance_matrix}
\end{equation}
with $\mathbf{1}$ a ($P \times 1$) vector containing ones. The phase variance terms $\sigma^{2}_{\phi}$ are removed from the equations by explicitly removing the mean phase from the individual phases through the substitution
\begin{eqnarray}
\phi & \to & \phi' = \phi - \langle \phi \rangle \quad \Rightarrow \nonumber \\
\mathbf{\Phi} & \to & \mathbf{\Phi}' = \left( \mathbb{I} - \frac{1}{P} \mathbf{1} \mathbf{1}^\mathrm{T}_{} \right) \mathbf{\Phi},
\label{eq:phi_subs}
\end{eqnarray}
where $\mathbb{I}$ is the ($P \times P$) identity matrix. Applying this substitution to the covariance matrix yields
\begin{eqnarray}
\mathbf{C}^{}_{\phi' \phi'} & = & \left( \mathbb{I} - \frac{1}{P} \mathbf{1} \mathbf{1}^\mathrm{T}_{} \right) \mathbf{C}^{}_{\phi \phi} \left( \mathbb{I} - \frac{1}{P} \mathbf{1} \mathbf{1}^\mathrm{T}_{} \right) \nonumber \\
 & = & \left( \mathbb{I} - \frac{1}{P} \mathbf{1} \mathbf{1}^\mathrm{T}_{} \right) \left[ - \frac{1}{2} \mathbf{D}^{}_{\phi \phi} \right] \left( \mathbb{I} - \frac{1}{P} \mathbf{1} \mathbf{1}^\mathrm{T}_{} \right),
\label{eq:cov_subs}
\end{eqnarray}
The $\sigma^{2}_{\phi}$ terms have dropped because of the properties
\begin{equation}
( \mathbb{I} - \frac{1}{P} \mathbf{1} \mathbf{1}^\mathrm{T}_{} ) \mathbf{1} = \mathbf{0}, \quad 
\mathbf{1}^\mathrm{T}_{} ( \mathbb{I} - \frac{1}{P} \mathbf{1} \mathbf{1}^\mathrm{T}_{} ) = \mathbf{0}
\label{eq:var_zero}
\end{equation}

For Kriging interpolation, a similar substitution is performed as in Equation~\ref{eq:phi_subs}:
\begin{equation}
\hat{\mathbf{\Phi}} \to \hat{\mathbf{\Phi}}' = \hat{\mathbf{\Phi}} - \left( \frac{1}{P} \hat{\mathbf{1}} \mathbf{1}^\mathrm{T}_{} \right) \mathbf{\Phi},
\label{eq:phi_hat_subs}
\end{equation}
with $\hat{\mathbf{1}}$ a ($\hat{P} \times 1$) vector containing ones. The covariance matrix between $\{\hat{\vec{p}}\}$ and $\{\vec{p}\}$ (Equation~\ref{eq:kriging}) now becomes
\begin{eqnarray}
\mathbf{C}^{}_{\hat{\phi}' \phi'} & = & \left( \left[ - \frac{1}{2} \mathbf{D}^{}_{\hat{\phi} \phi} \right] - \left( \frac{1}{P} \hat{\mathbf{1}} \mathbf{1}^\mathrm{T}_{} \right) \left[ - \frac{1}{2} \mathbf{D}^{}_{\phi \phi} \right] \right) \nonumber \\
& & \quad \left( \mathbb{I} - \frac{1}{P} \mathbf{1} \mathbf{1}^\mathrm{T}_{} \right)
\label{eq:kriging_subs}
\end{eqnarray}
where $\mathbf{D}^{}_{\hat{\phi} \phi}$ is the ($\hat{P} \times P$) structure matrix between $\{\hat{\vec{p}}\}$ and $\{\vec{p}\}$, calculated using Equations~\ref{eq:phase_structure_function} and \ref{eq:structure_matrix}.

\end{document}